\documentclass[aps,preprint]{revtex4}%
\usepackage{amsfonts}
\usepackage{amsmath}
\usepackage{amssymb}
\usepackage{graphicx}%
\begin{document}
\preprint{ }

\vspace*{1cm}

\begin{center}

{\Large Lattice chiral effective field theory with three-body interactions at
next-to-next-to-leading order}\vspace*{0.75cm}

{Evgeny Epelbaum$^{a,b}$, Hermann~Krebs$^{b,a}$, Dean~Lee$^{c,b}$,
Ulf-G.~Mei{\ss }ner$^{b,a,d}$}\vspace*{0.75cm}

$^{a}$\textit{Institut f\"{u}r Kernphysik (IKP-3) and J\"{u}lich Center for
Hadron Physics,}

\textit{Forschungszentrum J\"{u}lich, D-52425 J\"{u}lich, Germany }

$^{b}$\textit{Helmholtz-Institut f\"{u}r Strahlen- und Kernphysik (Theorie)}

\textit{and Bethe Center for Theoretical Physics, }\linebreak%
\textit{Universit\"{a}t Bonn, D-53115 Bonn, Germany }

$^{c}$\textit{Department of Physics, North Carolina State University, Raleigh,
NC 27695, USA}

$^{d}$\textit{Institute for Advanced Simulations (IAS), }

\textit{Forschungszentrum J\"{u}lich, D-52425 J\"{u}lich, Germany}%
\vspace*{0.75cm}

{\large Abstract}
\end{center}

We consider low-energy nucleons at next-to-next-to-leading order in lattice
chiral effective field theory. \ Three-body interactions first appear at this
order, and we discuss several methods for determining three-body interaction
coefficients on the lattice. \ We compute the energy of the triton and
low-energy neutron-deuteron scattering phase shifts in the spin-doublet and
spin-quartet channels using\ L\"{u}scher's finite volume method. \ In the
four-nucleon system we calculate the energy of the $\alpha$ particle using
auxiliary fields and projection Monte Carlo.\pagebreak

\section{Introduction}

We study low-energy nucleons on the lattice at next-to-next-to-leading order
in chiral effective field theory. \ In Weinberg's scheme
\cite{Weinberg:1990rz,Weinberg:1991um} counting orders in effective field
theory is equivalent to dimensional analysis for irreducible diagrams. \ The
expansion parameter is $Q$/$\Lambda$, where $Q$ is the low momentum scale
associated with external nucleon momenta or the pion mass, and $\Lambda$ is
the high momentum scale at which the effective theory breaks down. \ Terms at
next-to-next-to-leading order are of size $Q^{3}$/$\Lambda^{3}$, and
three-nucleon interactions first contribute at this order. \ In this work we
consider three-nucleon forces on the lattice for systems with three and four
nucleons. \ Our analysis continues a series of recent papers on lattice chiral
effective field theory for few- and many-nucleon systems. \ Previous studies
have considered dilute neutron matter and light nuclei using interactions at
leading order \cite{Lee:2004si,Borasoy:2006qn} and next-to-leading order
\cite{Borasoy:2007vi,Borasoy:2007vk,Epelbaum:2008vj}.

Our discussion is organized into three parts. \ The first part begins with an
overview of the effective potential for nucleons up to next-to-next-to-leading
order in chiral effective field theory. \ Reviews of chiral effective field
theory can be found in
Ref.~\cite{vanKolck:1999mw,Bedaque:2002mn,Epelbaum:2005pn,Epelbaum:2008ga}.
\ We discuss some simplifications that can be made at low cutoff momentum and
present lattice operators for each interaction. \ Nucleon-nucleon phase shifts
and the $S$-$D$ mixing angle are determined using a spherical wall method
\cite{Borasoy:2007vy} and used to set unknown operator coefficients. \ In the
second part we calculate the low-energy spectrum for three nucleons. \ We
compute the triton energy and determine neutron-deuteron phase shifts using
L\"{u}scher's finite volume method. \ These are used to constrain the two
unknown three-body operator coefficients. \ In the third and final part we
rewrite the lattice action in terms of auxiliary fields and use projection
Monte Carlo to calculate the energy of the $\alpha$ particle. \ This leads to
a discussion of alternative methods for fixing three-body operator
coefficients. \ We summarize our results and discuss possible extensions in
future work.

\section{Chiral effective field theory}

\subsection{Effective potential for two nucleons}

In the following $\vec{q}$ denotes the $t$-channel momentum transfer for
nucleon-nucleon scattering while $\vec{k}$ is the $u$-channel exchanged
momentum transfer. \ We assume exact isospin symmetry and neglect
electromagnetic interactions. \ At leading order (LO) in the Weinberg scheme
the two-nucleon effective potential consists of two independent contact terms
and instantaneous one-pion exchange (OPEP),%
\begin{equation}
V_{\text{LO}}=V^{(0)}+V^{\text{OPEP}}. \label{VLO}%
\end{equation}
The scattering between nucleons consists of contributions from direct and
exchange diagrams. \ Nevertheless for bookkeeping purposes we label the
interactions according to the tree-level scattering amplitude for
distinguishable nucleons. \ For two-nucleon interactions we label one nucleon
as type $A$ and the other nucleon as type $B$. \ In this notation the
amplitude for $V^{(0)}$ is%
\begin{equation}
\mathcal{A}\left[  V^{(0)}\right]  =C_{S}+C_{T}\left(  \vec{\sigma}_{A}%
\cdot\vec{\sigma}_{B}\right)  , \label{V0}%
\end{equation}
and the amplitude for $V^{\text{OPEP}}$ is%
\begin{equation}
\mathcal{A}\left[  V^{\text{OPEP}}\right]  =-\left(  \frac{g_{A}}{2f_{\pi}%
}\right)  ^{2}\boldsymbol\tau_{A}\cdot\boldsymbol\tau_{B}\frac{\left(
\vec{\sigma}_{A}\cdot\vec{q}\right)  \left(  \vec{\sigma}_{B}\cdot\vec
{q}\right)  }{q^{\,2}+m_{\pi}^{2}}. \label{VOPEP}%
\end{equation}
The vector arrow in $\vec{\sigma}$ signifies the three-vector index for spin.
\ The boldface for $\boldsymbol\tau$ signifies the three-vector index for
isospin. \ We take for our physical constants $m=938.92$~MeV as the nucleon
mass, $m_{\pi}=138.08$~MeV as the pion mass, $f_{\pi}=93$~MeV as the pion
decay constant, and $g_{A}=1.26$ as the nucleon axial charge.

At next-to-leading order (NLO) the two-nucleon effective potential contains
seven independent contact terms carrying two powers of momentum, corrections
to the two LO\ contact terms, and the leading contribution from the
instantaneous two-pion exchange potential (TPEP)
\cite{Ordonez:1992xp,Ordonez:1993tn,Ordonez:1996rz,Epelbaum:1998ka,Epelbaum:1999dj}%
,%
\begin{equation}
V_{\text{NLO}}=V_{\text{LO}}+\Delta V^{(0)}+V^{(2)}+V_{\text{NLO}%
}^{\text{TPEP}}. \label{VNLO}%
\end{equation}
The tree-level amplitudes for the contact interactions are%
\begin{equation}
\mathcal{A}\left[  \Delta V^{(0)}\right]  =\Delta C_{S}+\Delta C_{T}\left(
\vec{\sigma}_{A}\cdot\vec{\sigma}_{B}\right)  \label{dV0}%
\end{equation}
and%
\begin{align}
\mathcal{A}\left[  V^{(2)}\right]   &  =C_{1}q^{2}+C_{2}k^{2}+\left(
C_{3}q^{2}+C_{4}k^{2}\right)  \left(  \vec{\sigma}_{A}\cdot\vec{\sigma}%
_{B}\right)  +iC_{5}\frac{1}{2}\left(  \vec{\sigma}_{A}+\vec{\sigma}%
_{B}\right)  \cdot\left(  \vec{q}\times\vec{k}\right) \nonumber\\
&  +C_{6}\left(  \vec{\sigma}_{A}\cdot\vec{q}\right)  \left(  \vec{\sigma}%
_{B}\cdot\vec{q}\right)  +C_{7}\left(  \vec{\sigma}_{A}\cdot\vec{k}\right)
\left(  \vec{\sigma}_{B}\cdot\vec{k}\right)  . \label{V2}%
\end{align}
The amplitude for the NLO two-pion exchange potential is
\cite{Friar:1994,Kaiser:1997mw}%
\begin{align}
\mathcal{A}\left[  V_{\text{NLO}}^{\text{TPEP}}\right]   &  =-\frac
{\boldsymbol\tau_{A}\cdot\boldsymbol\tau_{B}}{384\pi^{2}f_{\pi}^{4}%
}L(q)\left[  4m_{\pi}^{2}\left(  5g_{A}^{4}-4g_{A}^{2}-1\right)  +q^{2}\left(
23g_{A}^{4}-10g_{A}^{2}-1\right)  +\frac{48g_{A}^{4}m_{\pi}^{4}}{4m_{\pi}%
^{2}+q^{2}}\right] \nonumber\\
&  -\frac{3g_{A}^{4}}{64\pi^{2}f_{\pi}^{4}}L(q)\left[  \left(  \vec{q}%
\cdot\vec{\sigma}_{A}\right)  \left(  \vec{q}\cdot\vec{\sigma}_{B}\right)
-q^{2}\left(  \vec{\sigma}_{A}\cdot\vec{\sigma}_{B}\right)  \right]  ,
\label{VTPEPNLO}%
\end{align}
where%
\begin{equation}
L(q)=\frac{1}{2q}\sqrt{4m_{\pi}^{2}+q^{2}}\ln\frac{\sqrt{4m_{\pi}^{2}+q^{2}%
}+q}{\sqrt{4m_{\pi}^{2}+q^{2}}-q}. \label{Lq}%
\end{equation}

At next-to-next-to-leading order (NNLO) there are no additional two-nucleon
contact interactions, but the two-pion exchange potential contains a
subleading contribution,%
\begin{align}
\mathcal{A}\left[  V_{\text{NNLO}}^{\text{TPEP}}\right]   &  =-\frac
{3g_{A}^{2}}{16\pi f_{\pi}^{4}}A(q)\left(  2m_{\pi}^{2}+q^{2}\right)  \left[
2m_{\pi}^{2}\left(  2c_{1}-c_{3}\right)  -c_{3}q^{2}\right] \nonumber\\
&  -\frac{g_{A}^{2}c_{4}\left(  \boldsymbol\tau_{A}\cdot\boldsymbol\tau
_{B}\right)  }{32\pi f_{\pi}^{4}}A(q)\left(  4m_{\pi}^{2}+q^{2}\right)
\left[  \left(  \vec{q}\cdot\vec{\sigma}_{A}\right)  \left(  \vec{q}\cdot
\vec{\sigma}_{B}\right)  -q^{2}\left(  \vec{\sigma}_{A}\cdot\vec{\sigma}%
_{B}\right)  \right]  ,
\end{align}
where%
\begin{equation}
A(q)=\frac{1}{2q}\arctan\frac{q}{2m_{\pi}}.
\end{equation}
The low-energy constants $c_{1},c_{3},c_{4}$ parameterize the coupling of the
nucleon to two pions. \ These constants have been determined from fits to
low-energy pion-nucleon scattering data \cite{Bernard:1995dp}, and in the
following we use the values $c_{1}=-0.81$~GeV$^{-1}$, $c_{3}=-4.7$~GeV$^{-1}$,
$c_{4}=3.4$~GeV$^{-1}$ \cite{Buettiker:1999ap}.

\subsection{Three-nucleon interactions}

\begin{figure}
[ptb]
\begin{center}
\includegraphics[
height=1.5869in,
width=2.4742in
]%
{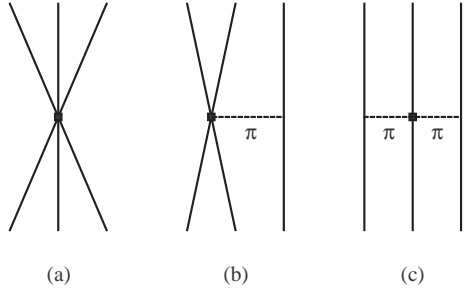}%
\caption{Three-nucleon forces at NNLO. \ Diagrams (a), (b), and (c) show the
contact potential, $V_{\text{contact}}^{(3N)}$, one-pion exchange potential
$V_{\text{OPE}}^{(3N)}$, and two-pion exchange potential $V_{\text{TPE}%
}^{(3N)}$.}%
\label{threebody}%
\end{center}
\end{figure}

A number of different phenomenological three-nucleon potentials have been
introduced in the literature
\cite{Fujita:1957zz,McKellar:1968a,Yang:1974zz,Coon:1974vc,Coon:1978gr,Coon:1981TM,Carlson:1983kq,Coelho:1984hk,Pudliner:1997ck}%
. \ Effective field theory provides a systematic method for estimating the
relative importance of three-body interaction terms. \ Few-nucleon forces in
chiral effective field theory beyond two nucleons were first introduced in
Ref.~\cite{Weinberg:1991um}. \ In Ref.~\cite{vanKolck:1994yi} it was shown
that three-nucleon interactions at NLO cancel and three-body effects first
appear at NNLO. \ The NNLO three-nucleon effective potential includes a pure
contact potential, $V_{\text{contact}}^{(3N)}$, one-pion exchange potential,
$V_{\text{OPE}}^{(3N)}$, and a two-pion exchange potential, $V_{\text{TPE}%
}^{(3N)}$,
\begin{equation}
V_{\text{NNLO}}^{(3N)}=V_{\text{contact}}^{(3N)}+V_{\text{OPE}}^{(3N)}%
+V_{\text{TPE}}^{(3N)}.
\end{equation}
The corresponding diagrams are shown in Fig.~\ref{threebody}.%

Similar to our bookkeeping notation for two-nucleon interactions, we write the
tree-level amplitude for three-nucleon interactions where the first nucleon is
type $A$, the second nucleon type $B$, and the third type $C$. \ We sum over
all permutations $P(A,B,C)$ of the labels, and $\vec{q}_{A}$, $\vec{q}_{B}$,
$\vec{q}_{C}$ are defined as the differences between final and initial momenta
for the respective nucleons. \ The amplitudes for $V_{\text{contact}}^{(3N)}$
and $V_{\text{OPE}}^{(3N)}$ are \cite{Friar:1998zt,Epelbaum:2002vt}%
\begin{equation}
\mathcal{A}\left[  V_{\text{contact}}^{(3N)}\right]  =\frac{1}{2}%
E\sum_{P(A,B,C)}\left(  \boldsymbol\tau_{A}\cdot\boldsymbol\tau_{B}\right)  ,
\label{contact_cont}%
\end{equation}%
\begin{equation}
\mathcal{A}\left[  V_{\text{OPE}}^{(3N)}\right]  =-\frac{g_{A}}{8f_{\pi}^{2}%
}D\sum_{P\left(  A,B,C\right)  }\frac{\vec{q}_{A}\cdot\vec{\sigma}_{A}}%
{q_{A}^{2}+m_{\pi}^{2}}\left(  \vec{q}_{A}\cdot\vec{\sigma}_{B}\right)
\left(  \boldsymbol\tau_{A}\cdot\boldsymbol\tau_{B}\right)  . \label{OPE_cont}%
\end{equation}
The coefficients $E$ and $D$ are both cutoff dependent. \ The coefficient $E$
determines the short distance interactions between three nucleons, while $D$
determines the pion coupling to two nucleons. \ Following the notation
introduced in Ref.~\cite{Epelbaum:2002vt}, we define dimensionless parameters
$c_{E}$ and $c_{D}$ such that
\begin{equation}
E=\frac{c_{E}}{f_{\pi}^{4}\Lambda_{\chi}},\quad D=\frac{c_{D}}{f_{\pi}%
^{2}\Lambda_{\chi}}\text{,}%
\end{equation}
where $\Lambda_{\chi}\simeq m_{\rho}$. \ We take $\Lambda_{\chi}=700$~MeV.

For convenience we separately label three parts of the two-pion exchange
potential$,$%
\begin{equation}
V_{\text{TPE}}^{(3N)}=V_{\text{TPE1}}^{(3N)}+V_{\text{TPE2}}^{(3N)}%
+V_{\text{TPE3}}^{(3N)}.
\end{equation}
The corresponding amplitudes are%
\begin{equation}
\mathcal{A}\left[  V_{\text{TPE1}}^{(3N)}\right]  =\frac{c_{3}}{f_{\pi}^{2}%
}\left(  \frac{g_{A}}{2f_{\pi}}\right)  ^{2}\sum_{P\left(  A,B,C\right)
}\frac{\left(  \vec{q}_{A}\cdot\vec{\sigma}_{A}\right)  \left(  \vec{q}%
_{B}\cdot\vec{\sigma}_{B}\right)  }{\left(  q_{A}^{2}+m_{\pi}^{2}\right)
\left(  q_{B}^{2}+m_{\pi}^{2}\right)  }\left(  \vec{q}_{A}\cdot\vec{q}%
_{B}\right)  \left(  \boldsymbol\tau_{A}\cdot\boldsymbol\tau_{B}\right)  ,
\label{TPE1_cont}%
\end{equation}%
\begin{equation}
\mathcal{A}\left[  V_{\text{TPE2}}^{(3N)}\right]  =-\frac{2c_{1}m_{\pi}^{2}%
}{f_{\pi}^{2}}\left(  \frac{g_{A}}{2f_{\pi}}\right)  ^{2}\sum_{P\left(
A,B,C\right)  }\frac{\left(  \vec{q}_{A}\cdot\vec{\sigma}_{A}\right)  \left(
\vec{q}_{B}\cdot\vec{\sigma}_{B}\right)  }{\left(  q_{A}^{2}+m_{\pi}%
^{2}\right)  \left(  q_{B}^{2}+m_{\pi}^{2}\right)  }\left(  \boldsymbol\tau
_{A}\cdot\boldsymbol\tau_{B}\right)  , \label{TPE2_cont}%
\end{equation}%
\begin{align}
\mathcal{A}\left[  V_{\text{TPE3}}^{(3N)}\right]   &  =\frac{c_{4}}{2f_{\pi
}^{2}}\left(  \frac{g_{A}}{2f_{\pi}}\right)  ^{2}\nonumber\\
&  \times\sum_{P\left(  A,B,C\right)  }\frac{\left(  \vec{q}_{A}\cdot
\vec{\sigma}_{A}\right)  \left(  \vec{q}_{B}\cdot\vec{\sigma}_{B}\right)
}{\left(  q_{A}^{2}+m_{\pi}^{2}\right)  \left(  q_{B}^{2}+m_{\pi}^{2}\right)
}\left[  \left(  \vec{q}_{A}\times\vec{q}_{B}\right)  \cdot\vec{\sigma}%
_{C}\right]  \left[  \left(  \boldsymbol\tau_{A}\times\boldsymbol\tau
_{B}\right)  \cdot\boldsymbol\tau_{C}\right]  . \label{TPE3_cont}%
\end{align}

\subsection{Simplified form at low cutoff momentum}

A number of alternatives to Weinberg's power counting have recently been
discussed in the literature
\cite{Beane:2001bc,Nogga:2005hy,Birse:2005um,Birse:2007sx}. \ However at low
cutoff momentum the advantages of these alternative schemes are numerically
small \cite{Epelbaum:2006pt}. \ In this study we use spatial lattice spacing
$a=(100$~MeV$)^{-1}$, corresponding with cutoff momentum $\Lambda=314$~MeV
$\approx2.3m_{\pi}$. \ Our choice of low cutoff scale avoids numerical
problems in Monte Carlo simulations due to spurious deeply-bound states and
large sign/phase oscillations. \ 

In the following lattice calculations we use Weinberg's power counting with
some additional simplifications made possible by the low cutoff momentum.
\ For nearly all $\left\vert q\right\vert <\Lambda$ we can expand the NLO
two-pion exchange potential in powers of $q^{2}/(4m_{\pi}^{2}),$%
\begin{align}
\mathcal{A}\left[  V_{\text{NLO}}^{\text{TPEP}}\right]   &  =-\frac
{\boldsymbol\tau_{A}\cdot\boldsymbol\tau_{B}}{384\pi^{2}f_{\pi}^{4}}\left[
4m_{\pi}^{2}\left(  8g_{A}^{4}-4g_{A}^{2}-1\right)  +\frac{2}{3}q^{2}\left(
34g_{A}^{4}-17g_{A}^{2}-2\right)  +m_{\pi}^{2}O\left(  \left(  \tfrac{q^{2}%
}{4m_{\pi}^{2}}\right)  ^{2}\right)  \right] \nonumber\\
&  -\frac{3g_{A}^{4}}{64\pi^{2}f_{\pi}^{4}}\left[  \left(  \vec{q}\cdot
\vec{\sigma}_{A}\right)  \left(  \vec{q}\cdot\vec{\sigma}_{B}\right)
-q^{2}\left(  \vec{\sigma}_{A}\cdot\vec{\sigma}_{B}\right)  \right]  \left[
1+O\left(  \tfrac{q^{2}}{4m_{\pi}^{2}}\right)  \right]  . \label{localTPEP}%
\end{align}
This expansion fails to converge only for values of $q$ near the cutoff scale
$\Lambda$ $\approx2.3m_{\pi}$, where the effective theory is already
problematic due to large cutoff effects. \ From a practical viewpoint there is
no advantage in retaining the full non-local structure of $V_{\text{NLO}%
}^{\text{TPEP}}$ at this lattice spacing. \ Instead we simply use%
\begin{equation}
V_{\text{LO}}=V^{(0)}+V^{\text{OPEP}},
\end{equation}%
\begin{equation}
V_{\text{NLO}}=V_{\text{LO}}+\Delta V^{(0)}+V^{(2)},
\end{equation}
where the terms in Eq.~(\ref{localTPEP}) with up to two powers of $q$ are
absorbed as a redefinition of the coefficients $\Delta V^{(0)}$ and $V^{(2)}$.

Similarly we can expand the NNLO two-pion exchange potential,%
\begin{align}
\mathcal{A}\left[  V_{\text{NNLO}}^{\text{TPEP}}\right]   &  =-\frac
{3g_{A}^{2}m_{\pi}}{16\pi f_{\pi}^{4}}\left[  m_{\pi}^{2}\left(  2c_{1}%
-c_{3}\right)  +q^{2}\left(  \frac{5}{6}c_{1}-\frac{11}{12}c_{3}\right)
+m_{\pi}^{2}O\left(  \left(  \tfrac{q^{2}}{4m_{\pi}^{2}}\right)  ^{2}\right)
\right] \nonumber\\
&  -\frac{g_{A}^{2}c_{4}\left(  \boldsymbol\tau_{A}\cdot\boldsymbol\tau
_{B}\right)  m_{\pi}}{32\pi f_{\pi}^{4}}\left[  \left(  \vec{q}\cdot
\vec{\sigma}_{A}\right)  \left(  \vec{q}\cdot\vec{\sigma}_{B}\right)
-q^{2}\left(  \vec{\sigma}_{A}\cdot\vec{\sigma}_{B}\right)  \right]  \left[
1+O\left(  \tfrac{q^{2}}{4m_{\pi}^{2}}\right)  \right]  .
\end{align}
The terms with two powers of $q$ were already included at NLO, and so there
are no additional terms in the two-nucleon potential at NNLO. \ In our low
cutoff scheme the only new contributions at NNLO are due to three-nucleon
interactions,%
\begin{equation}
V_{\text{NNLO}}=V_{\text{NLO}}+V_{\text{NNLO}}^{(3N)}.
\end{equation}

\section{Lattice interactions at LO\ and NLO}

\subsection{Transfer matrix at LO}

\begin{figure}
[ptb]
\begin{center}
\includegraphics[
height=1.6362in,
width=2.8842in
]%
{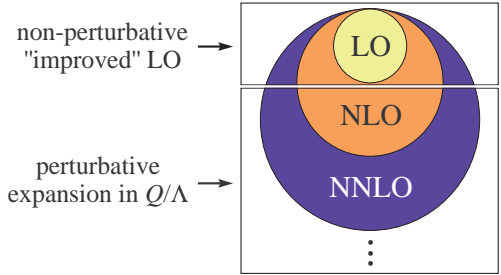}%
\caption{The \textquotedblleft improved\textquotedblright\ LO action is
iterated non-perturbatively while the remaining higher-order interactions are
treated using perturbation theory.}%
\label{eftorders}%
\end{center}
\end{figure}

In our Euclidean-time lattice formalism the transfer matrix operator is the
normal-ordered exponential of the lattice Hamiltonian, $\colon\exp(-H\Delta
t)\colon$, where $\Delta t$ equals one temporal lattice spacing, $a_{t}$. \ At
leading order we use the LO$_{2}$ transfer matrix with
Gaussian-smeared\ interactions~\cite{Borasoy:2006qn,Borasoy:2007vi,Borasoy:2007vk}%
. \ Since we consider only one action, we drop the \textquotedblleft%
$2$\textquotedblright\ subscript on LO$_{2}$. \ The transfer matrix operator
is
\begin{align}
M_{\text{LO}}  &  =\colon\exp\left\{  -H_{\text{free}}\alpha_{t}-\frac
{\alpha_{t}}{2L^{3}}\sum_{\vec{q}}f(\vec{q})\left[  C\rho^{a^{\dag},a}(\vec
{q})\rho^{a^{\dag},a}(-\vec{q})+C_{I^{2}}\sum_{I}\rho_{I}^{a^{\dag},a}(\vec
{q})\rho_{I}^{a^{\dag},a}(-\vec{q})\right]  \right. \nonumber\\
&  +\left.  \frac{g_{A}^{2}\alpha_{t}^{2}}{8f_{\pi}^{2}q_{\pi}}\sum
_{\substack{S_{1},S_{2},I}}\sum_{\vec{n}_{1},\vec{n}_{2}}G_{S_{1}S_{2}}%
(\vec{n}_{1}-\vec{n}_{2})\rho_{S_{1},I}^{a^{\dag},a}(\vec{n}_{1})\rho
_{S_{2},I}^{a^{\dag},a}(\vec{n}_{2})\right\}  \colon. \label{LO}%
\end{align}
The momentum-dependent coefficient function $f(\vec{q})$ is given by
\begin{equation}
f(\vec{q})=f_{0}^{-1}\exp\left[  -b%
{\displaystyle\sum\limits_{l}}
\left(  1-\cos q_{l}\right)  \right]  ,
\end{equation}
where%
\begin{equation}
f_{0}=\frac{1}{L^{3}}\sum_{\vec{q}}\exp\left[  -b%
{\displaystyle\sum\limits_{l}}
\left(  1-\cos q_{l}\right)  \right]  .
\end{equation}
Our lattice notation is defined in the Appendix. \ The densities
$\rho^{a^{\dagger},a}$ and $\rho_{I}^{a^{\dagger},a}$ and spin-dependent
one-pion exchange potential $G_{S_{1}S_{2}}$ are also defined in the Appendix.
\ We use the value $b=0.6$, which gives approximately the correct effective
range for the two $S$-wave channels when $C$ and $C_{I^{2}}$ are tuned to the
physical scattering lengths. \ $C$ is the coefficient of the Wigner
SU(4)-invariant contact interaction \cite{Wigner:1937},\ and $C_{I^{2}}$ is
the coefficient of the isospin-dependent contact interaction. \ In terms of
coefficients for the isospin-singlet and triplet channels,
\begin{equation}
C=\left(  3C^{I=1}+C^{I=0}\right)  /4, \label{C_coeff}%
\end{equation}%
\begin{equation}
C_{I^{2}}=\left(  C^{I=1}-C^{I=0}\right)  /4. \label{C_I2_coeff}%
\end{equation}
This \textquotedblleft improved\textquotedblright\ leading-order\ action is
treated non-perturbatively while higher-order interactions are included as a
perturbative expansion in powers of $Q/\Lambda$. \ This is sketched in
Fig.~\ref{eftorders}.%

In pionless effective field theory the three-nucleon contact interaction is
included at leading order \cite{Bedaque:1998kg,Bedaque:1998km,Bedaque:1999ve}.
\ This is needed to stabilize the three-nucleon system in the limit of
zero-range interactions \cite{Thomas:1935}. \ In this study we use chiral
effective field theory where the interactions have nonzero range. \ However
given our coarse lattice spacing, we may find that the three-body contact
interaction is numerically large and requires non-perturbative treatment.
\ Non-perturbative three-body contact interactions on the lattice have been
discussed in the literature \cite{Chen:2004rq,Borasoy:2005yc,Lee:2008fa}.
\ However for the lattice calculations presented here we choose a different
approach to address the same problem.

We use the fact that the three-nucleon interaction depends on both the spatial
lattice spacing, $a$, and the temporal lattice spacing, $a_{t}$. \ The
temporal lattice spacing regulates the transfer matrix element when the
interaction potential energy exceeds $a_{t}^{-1}$. \ As a result it can affect
the magnitude and sign of the three-body contact interaction \cite{Lee:2005xy}%
. \ With the spatial lattice spacing held fixed, we dial the temporal lattice
spacing to a value where the three-nucleon interaction is numerically small.
\ This involves a calculation of the spectrum of the three-nucleon system.
\ In the following calculations we use the lattice spacings $a=(100$%
~MeV$)^{-1}$ and $a_{t}=(150$~MeV$)^{-1}$ and show that the strength of the
three-nucleon contact interaction is small enough to be treated using
perturbation theory. \ For these lattice spacings we find leading-order
coefficients $C^{I=0}=-5.105\times10^{-5}$~MeV$^{-2}$ and $C^{I=1}%
=-3.507\times10^{-5}$~MeV$^{-2}$ when tuned to the physical $S$-wave
scattering lengths$.$

\subsection{Transfer matrix at NLO}

At next-to-leading order the lattice transfer matrix is%
\begin{align}
M_{\text{NLO}}  &  =M_{\text{LO}}-\left.  \alpha_{t}\colon\left[  \Delta
V+\Delta V_{I^{2}}+V_{q^{2}}+V_{I^{2},q^{2}}+V_{S^{2},q^{2}}\right.  \right.
\nonumber\\
&  \left.  \qquad\qquad\qquad\qquad+V_{S^{2},I^{2},q^{2}}+V_{(q\cdot S)^{2}%
}+V_{I^{2},(q\cdot S)^{2}}+V_{(iq\times S)\cdot k}^{I=1}\right]  M_{\text{LO}%
}\colon\text{.}%
\end{align}
The corrections to the leading-order contact interactions are%
\begin{equation}
\Delta V=\frac{1}{2}\Delta C:\sum\limits_{\vec{n}}\rho^{a^{\dagger},a}(\vec
{n})\rho^{a^{\dagger},a}(\vec{n}):,
\end{equation}%
\begin{equation}
\Delta V_{I^{2}}=\frac{1}{2}\Delta C_{I^{2}}:\sum\limits_{\vec{n},I}\rho
_{I}^{a^{\dagger},a}(\vec{n})\rho_{I}^{a^{\dagger},a}(\vec{n}):,
\end{equation}
and the seven independent contact interactions with two derivatives are%
\begin{equation}
V_{q^{2}}=-\frac{1}{2}C_{q^{2}}:\sum\limits_{\vec{n},l}\rho^{a^{\dagger}%
,a}(\vec{n})\triangledown_{l}^{2}\rho^{a^{\dagger},a}(\vec{n}):,
\end{equation}%
\begin{equation}
V_{I^{2},q^{2}}=-\frac{1}{2}C_{I^{2},q^{2}}:\sum\limits_{\vec{n},I,l}\rho
_{I}^{a^{\dagger},a}(\vec{n})\triangledown_{l}^{2}\rho_{I}^{a^{\dagger}%
,a}(\vec{n}):,
\end{equation}%
\begin{equation}
V_{S^{2},q^{2}}=-\frac{1}{2}C_{S^{2},q^{2}}:\sum\limits_{\vec{n},S,l}\rho
_{S}^{a^{\dagger},a}(\vec{n})\triangledown_{l}^{2}\rho_{S}^{a^{\dagger}%
,a}(\vec{n}):,
\end{equation}%
\begin{equation}
V_{S^{2},I^{2},q^{2}}=-\frac{1}{2}C_{S^{2},I^{2},q^{2}}:\sum\limits_{\vec
{n},S,I,l}\rho_{S,I}^{a^{\dagger},a}(\vec{n})\triangledown_{l}^{2}\rho
_{S,I}^{a^{\dagger},a}(\vec{n}):,
\end{equation}%
\begin{equation}
V_{(q\cdot S)^{2}}=\frac{1}{2}C_{(q\cdot S)^{2}}:\sum\limits_{\vec{n}}%
\sum\limits_{S}\Delta_{S}\rho_{S}^{a^{\dagger},a}(\vec{n})\sum
\limits_{S^{\prime}}\Delta_{S^{\prime}}\rho_{S^{\prime}}^{a^{\dagger},a}%
(\vec{n}):,
\end{equation}%
\begin{equation}
V_{I^{2},(q\cdot S)^{2}}=\frac{1}{2}C_{I^{2},(q\cdot S)^{2}}:\sum
\limits_{\vec{n},I}\sum\limits_{S}\Delta_{S}\rho_{S,I}^{a^{\dagger},a}(\vec
{n})\sum\limits_{S^{\prime}}\Delta_{S^{\prime}}\rho_{S^{\prime},I}%
^{a^{\dagger},a}(\vec{n}):,
\end{equation}%
\begin{align}
V_{(iq\times S)\cdot k}^{I=1}  &  =-\frac{i}{2}C_{(iq\times S)\cdot k}%
^{I=1}\left\{  \frac{3}{4}:\sum\limits_{\vec{n},l,S,l^{\prime}}\varepsilon
_{l,S,l^{\prime}}\left[  \Pi_{l}^{a^{\dagger},a}(\vec{n})\Delta_{l^{\prime}%
}\rho_{S}^{a^{\dagger},a}(\vec{n})+\Pi_{l,S}^{a^{\dagger},a}(\vec{n}%
)\Delta_{l^{\prime}}\rho^{a^{\dagger},a}(\vec{n})\right]  :\right. \nonumber\\
&  +\left.  \frac{1}{4}:\sum\limits_{\vec{n},l,S,l^{\prime},I}\varepsilon
_{l,S,l^{\prime}}\left[  \Pi_{l,I}^{a^{\dagger},a}(\vec{n})\Delta_{l^{\prime}%
}\rho_{S,I}^{a^{\dagger},a}(\vec{n})+\Pi_{l,S,I}^{a^{\dagger},a}(\vec
{n})\Delta_{l^{\prime}}\rho_{I}^{a^{\dagger},a}(\vec{n})\right]  :\right\}  .
\end{align}
The densities, current densities, and symbols $\Delta_{l}$ and $\triangledown
_{l}^{2}$, are defined in the Appendix.

The $V_{(iq\times S)\cdot k}^{I=1}$ term is designed to eliminate lattice
artifacts in the spin-triplet even-parity channels. \ This is done by
projecting onto the isospin-triplet state,%
\begin{equation}
V_{(iq\times S)\cdot k}^{I=1}=V_{(iq\times S)\cdot k}+V_{I^{2},(iq\times
S)\cdot k},
\end{equation}
where%
\begin{equation}
V_{(iq\times S)\cdot k}=-\frac{i}{2}C_{(iq\times S)\cdot k}:\sum
\limits_{\vec{n},l,S,l^{\prime}}\varepsilon_{l,S,l^{\prime}}\left[  \Pi
_{l}^{a^{\dagger},a}(\vec{n})\Delta_{l^{\prime}}\rho_{S}^{a^{\dagger},a}%
(\vec{n})+\Pi_{l,S}^{a^{\dagger},a}(\vec{n})\Delta_{l^{\prime}}\rho
^{a^{\dagger},a}(\vec{n})\right]  :,
\end{equation}%
\begin{equation}
V_{I^{2},(iq\times S)\cdot k}=-\frac{i}{2}C_{I^{2},(iq\times S)\cdot k}%
:\sum\limits_{\vec{n},l,S,l^{\prime},I}\varepsilon_{l,S,l^{\prime}}\left[
\Pi_{l,I}^{a^{\dagger},a}(\vec{n})\Delta_{l^{\prime}}\rho_{S,I}^{a^{\dagger
},a}(\vec{n})+\Pi_{l,S,I}^{a^{\dagger},a}(\vec{n})\Delta_{l^{\prime}}\rho
_{I}^{a^{\dagger},a}(\vec{n})\right]  :,
\end{equation}
and%
\begin{equation}
C_{(iq\times S)\cdot k}=\frac{3}{4}C_{(iq\times S)\cdot k}^{I=1},
\end{equation}%
\begin{equation}
C_{I^{2},(iq\times S)\cdot k}=\frac{1}{4}C_{(iq\times S)\cdot k}^{I=1}.
\end{equation}

We measure phase shifts on the lattice by imposing a spherical wall boundary
on the relative separation between two nucleons at some chosen radius. \ From
the properties of the spherical standing waves we determine scattering phase
shifts and mixing angles \cite{Borasoy:2007vy}. \ The scattering results are
nearly identical with the LO$_{2}$ data at lattice spacings $a=(100$%
~MeV$)^{-1}$ and $a_{t}=(70$~MeV$)^{-1}$ presented in
Ref.~\cite{Borasoy:2007vi}. \ The values for the next-to-leading order
coefficients are shown in Table~\ref{NLOcoeff}. \ These values are similar to
the NLO$_{2}$ coefficients in Table~III of Ref.~\cite{Borasoy:2007vi}, though
there are some differences due to the change in temporal lattice spacing.

\begin{table}[tb]
\caption{Results for the NLO operator coefficients}%
\label{NLOcoeff}%
\begin{tabular}
[c]{||c|c||}\hline\hline
Coefficient & Value\\\hline
$\Delta C$ [MeV$^{-2}$] & $4.08\times10^{-6}$\\\hline
$\Delta C_{I^{2}}$ [MeV$^{-2}$] & $5.92\times10^{-6}$\\\hline
$C_{q^{2}}$ [MeV$^{-4}$] & $-1.31\times10^{-9}$\\\hline
$C_{I^{2},q^{2}}$ [MeV$^{-4}$] & $-3.26\times10^{-10}$\\\hline
$C_{S^{2},q^{2}}$ [MeV$^{-4}$] & $-1.53\times10^{-10}$\\\hline
$C_{S^{2},I^{2},q^{2}}$ [MeV$^{-4}$] & $-2.64\times10^{-10}$\\\hline
$C_{(q\cdot S)^{2}}$ [MeV$^{-4}$] & $-1.92\times10^{-10}$\\\hline
$C_{I^{2},(q\cdot S)^{2}}$ [MeV$^{-4}$] & $9.20\times10^{-12}$\\\hline
$C_{(iq\times S)\cdot k}^{I=1}$ [MeV$^{-4}$] & $1.11\times10^{-10}%
$\\\hline\hline
\end{tabular}
\end{table}

\section{Three-nucleon interactions at NNLO}

At next-to-next-to-leading order the transfer matrix is%
\begin{equation}
M_{\text{NNLO}}=M_{\text{NLO}}-\left.  \alpha_{t}\colon\right.  \left[
V_{\text{contact}}^{(3N)}+V_{\text{OPE}}^{(3N)}+V_{\text{TPE1}}^{(3N)}%
+V_{\text{TPE2}}^{(3N)}+V_{\text{TPE3}}^{(3N)}\right]  M_{\text{LO}}:.
\end{equation}
From the constraints of isospin symmetry, spin symmetry, and Fermi statistics,
there is only one independent three-nucleon contact interaction
\cite{Bedaque:1999ve,Epelbaum:2002vt}. \ In Eq.~(\ref{contact_cont}) we wrote
this as a $\boldsymbol\tau_{A}\cdot\boldsymbol\tau_{B}$ interaction over all
permutations of the labels $A,B,C$. \ For our lattice action we choose to
write the contact interaction $V_{\text{contact}}^{(3N)}$ as a product of
total nucleon densities,%
\begin{equation}
V_{\text{contact}}^{(3N)}=\frac{1}{6}D_{\text{contact}}:\sum_{\vec{n}}\left[
\rho^{a^{\dagger},a}(\vec{n})\right]  ^{3}:\text{.}%
\end{equation}
The one-pion exchange potential $V_{\text{OPE}}^{(3N)}$ can be written as%
\begin{equation}
V_{\text{OPE}}^{(3N)}=-D_{\text{OPE}}\frac{g_{A}\alpha_{t}}{2f_{\pi}q_{\pi}%
}\sum_{\vec{n},S,I}\sum_{\vec{n}^{\prime},S^{\prime}}\left\langle
\Delta_{S^{\prime}}\pi_{I}^{\prime}(\vec{n}^{\prime},n_{t})\Delta_{S}\pi
_{I}^{\prime}(\vec{n},n_{t})\right\rangle :\rho_{S^{\prime},I}^{a^{\dag}%
,a}(\vec{n}^{\prime})\rho_{S,I}^{a^{\dag},a}(\vec{n})\rho^{a^{\dag},a}(\vec
{n}):\text{.}%
\end{equation}
The three two-pion exchange terms $V_{\text{TPE1}}^{(3N)},$ $V_{\text{TPE2}%
}^{(3N)},$ $V_{\text{TPE3}}^{(3N)}$ are%
\begin{align}
V_{\text{TPE1}}^{(3N)}  &  =D_{\text{TPE1}}\frac{g_{A}^{2}\alpha_{t}^{2}%
}{4f_{\pi}^{2}q_{\pi}^{2}}\sum_{\vec{n},S,I}\sum_{\vec{n}^{\prime},S^{\prime}%
}\sum_{\vec{n}^{\prime\prime},S^{\prime\prime}}\left[
\begin{array}
[c]{c}%
\!\\
\!
\end{array}
\left\langle \Delta_{S^{\prime}}\pi_{I}^{\prime}(\vec{n}^{\prime},n_{t}%
)\Delta_{S}\pi_{I}^{\prime}(\vec{n},n_{t})\right\rangle \right. \nonumber\\
&  \times\left.  \left\langle \Delta_{S^{\prime\prime}}\pi_{I}^{\prime}%
(\vec{n}^{\prime\prime},n_{t})\Delta_{S}\pi_{I}^{\prime}(\vec{n}%
,n_{t})\right\rangle :\rho_{S^{\prime},I}^{a^{\dag},a}(\vec{n}^{\prime}%
)\rho_{S^{\prime\prime},I}^{a^{\dag},a}(\vec{n}^{\prime\prime})\rho^{a^{\dag
},a}(\vec{n}):%
\begin{array}
[c]{c}%
\!\\
\!
\end{array}
\right]  \text{,}%
\end{align}

\begin{align}
V_{\text{TPE2}}^{(3N)}  &  =D_{\text{TPE2}}m_{\pi}^{2}\frac{g_{A}^{2}%
\alpha_{t}^{2}}{4f_{\pi}^{2}q_{\pi}^{2}}\sum_{\vec{n},I}\sum_{\vec{n}^{\prime
},S^{\prime}}\sum_{\vec{n}^{\prime\prime},S^{\prime\prime}}\left[
\begin{array}
[c]{c}%
\!\\
\!
\end{array}
\left\langle \Delta_{S^{\prime}}\pi_{I}^{\prime}(\vec{n}^{\prime}%
,n_{t})\square\pi_{I}^{\prime}(\vec{n},n_{t})\right\rangle \right. \nonumber\\
&  \times\left.  \left\langle \Delta_{S^{\prime\prime}}\pi_{I}^{\prime}%
(\vec{n}^{\prime\prime},n_{t})\square\pi_{I}^{\prime}(\vec{n},n_{t}%
)\right\rangle :\rho_{S^{\prime},I}^{a^{\dag},a}(\vec{n}^{\prime}%
)\rho_{S^{\prime\prime},I}^{a^{\dag},a}(\vec{n}^{\prime\prime})\rho^{a^{\dag
},a}(\vec{n}):%
\begin{array}
[c]{c}%
\!\\
\!
\end{array}
\right]  ,
\end{align}%
\begin{align}
V_{\text{TPE3}}^{(3N)}  &  =D_{\text{TPE3}}\frac{g_{A}^{2}\alpha_{t}^{2}%
}{4f_{\pi}^{2}q_{\pi}^{2}}\sum_{\vec{n},S_{1},S_{2},S_{3}}\sum_{I_{1}%
,I_{2},I_{3}}\sum_{\vec{n}^{\prime},S^{\prime}}\sum_{\vec{n}^{\prime\prime
},S^{\prime\prime}}\left[
\begin{array}
[c]{c}%
\!\\
\!
\end{array}
\right. \nonumber\\
&  \times\left\langle \Delta_{S^{\prime}}\pi_{I_{1}}^{\prime}(\vec{n}^{\prime
},n_{t})\Delta_{S_{1}}\pi_{I_{1}}^{\prime}(\vec{n},n_{t})\right\rangle
\left\langle \Delta_{S^{\prime\prime}}\pi_{I_{2}}^{\prime}(\vec{n}%
^{\prime\prime},n_{t})\Delta_{S_{2}}\pi_{I_{2}}^{\prime}(\vec{n}%
,n_{t})\right\rangle \nonumber\\
&  \times\left.  \varepsilon_{S_{1},S_{2},S_{3}}\varepsilon_{I_{1},I_{2}%
,I_{3}}:\rho_{S^{\prime},I_{1}}^{a^{\dag},a}(\vec{n}^{\prime})\rho
_{S^{\prime\prime},I_{2}}^{a^{\dag},a}(\vec{n}^{\prime\prime})\rho
_{S_{3},I_{3}}^{a^{\dag},a}(\vec{n}):%
\begin{array}
[c]{c}%
\!\\
\!
\end{array}
\right]  \text{.}%
\end{align}
Definitions for the $\square$ symbol and the two-point pion correlation
functions are given in the Appendix.

In the continuum limit the tree-level scattering amplitudes are%
\begin{equation}
\mathcal{A}\left[  V_{\text{contact}}^{(3N)}\right]  =D_{\text{contact}%
}\text{,}%
\end{equation}%
\begin{equation}
\mathcal{A}\left[  V_{\text{OPE}}^{(3N)}\right]  =-D_{\text{OPE}}\frac{g_{A}%
}{2f_{\pi}}\sum_{P\left(  A,B,C\right)  }\frac{\vec{q}_{A}\cdot\vec{\sigma
}_{A}}{q_{A}^{2}+m_{\pi}^{2}}\left(  \vec{q}_{A}\cdot\vec{\sigma}_{B}\right)
\left(  \boldsymbol\tau_{A}\cdot\boldsymbol\tau_{B}\right)  ,
\end{equation}%
\begin{equation}
\mathcal{A}\left[  V_{\text{TPE1}}^{(3N)}\right]  =D_{\text{TPE1}}\frac
{g_{A}^{2}}{4f_{\pi}^{2}}\sum_{P\left(  A,B,C\right)  }\frac{\left(  \vec
{q}_{A}\cdot\vec{\sigma}_{A}\right)  \left(  \vec{q}_{B}\cdot\vec{\sigma}%
_{B}\right)  }{\left(  q_{A}^{2}+m_{\pi}^{2}\right)  \left(  q_{B}^{2}+m_{\pi
}^{2}\right)  }\left(  \vec{q}_{A}\cdot\vec{q}_{B}\right)  \left(
\boldsymbol\tau_{A}\cdot\boldsymbol\tau_{B}\right)  ,
\end{equation}%
\begin{equation}
\mathcal{A}\left[  V_{\text{TPE2}}^{(3N)}\right]  =D_{\text{TPE2}}m_{\pi}%
^{2}\frac{g_{A}^{2}}{4f_{\pi}^{2}}\sum_{P\left(  A,B,C\right)  }\frac{\left(
\vec{q}_{A}\cdot\vec{\sigma}_{A}\right)  \left(  \vec{q}_{B}\cdot\vec{\sigma
}_{B}\right)  }{\left(  q_{A}^{2}+m_{\pi}^{2}\right)  \left(  q_{B}^{2}%
+m_{\pi}^{2}\right)  }\left(  \boldsymbol\tau_{A}\cdot\boldsymbol\tau
_{B}\right)  ,
\end{equation}

\begin{equation}
\mathcal{A}\left[  V_{\text{TPE3}}^{(3N)}\right]  =D_{\text{TPE3}}\frac
{g_{A}^{2}}{4f_{\pi}^{2}}\sum_{P\left(  A,B,C\right)  }\frac{\left(  \vec
{q}_{A}\cdot\vec{\sigma}_{A}\right)  \left(  \vec{q}_{B}\cdot\vec{\sigma}%
_{B}\right)  }{\left(  q_{A}^{2}+m_{\pi}^{2}\right)  \left(  q_{B}^{2}+m_{\pi
}^{2}\right)  }\left[  \left(  \vec{q}_{A}\times\vec{q}_{B}\right)  \cdot
\vec{\sigma}_{C}\right]  \left[  \left(  \boldsymbol\tau_{A}\times
\boldsymbol\tau_{B}\right)  \cdot\boldsymbol\tau_{C}\right]  .
\end{equation}
Comparing these with Eq.~(\ref{contact_cont}-\ref{TPE3_cont}), we have%
\begin{equation}
D_{\text{contact}}=-3E=-\frac{3c_{E}}{f_{\pi}^{4}\Lambda_{\chi}},\qquad
D_{\text{OPE}}=\frac{D}{4f_{\pi}}=\frac{c_{D}}{4f_{\pi}^{3}\Lambda_{\chi}},
\end{equation}%
\begin{equation}
D_{\text{TPE1}}=\frac{c_{3}}{f_{\pi}^{2}},\qquad D_{\text{TPE2}}=-\frac
{2c_{1}}{f_{\pi}^{2}},\qquad D_{\text{TPE3}}=\frac{c_{4}}{2f_{\pi}^{2}}.
\end{equation}

\subsection{Triton energy}

\begin{figure}
[ptb]
\begin{center}
\includegraphics[
height=2.4241in,
width=2.6688in
]%
{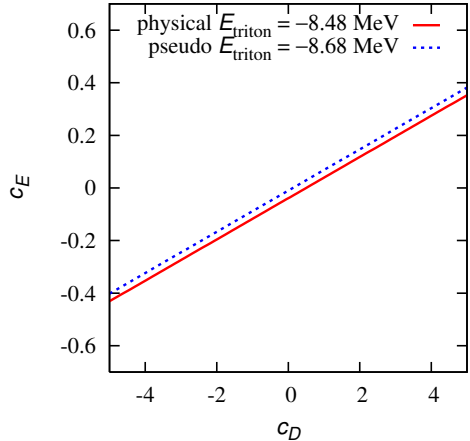}%
\caption{Plot of $c_{E}$ versus $c_{D}$ when constrained by the triton energy
at infinite volume. \ The solid line shows data matched to the physical triton
energy, $-8.48$ MeV. \ The dotted line shows results fitted to the pseudo
triton energy, $-8.68$ MeV, estimated by replacing $nn$ forces with $np$
forces.}%
\label{ce_vs_cd}%
\end{center}
\end{figure}

\begin{figure}
[ptbptb]
\begin{center}
\includegraphics[
height=3.0277in,
width=2.5425in
]%
{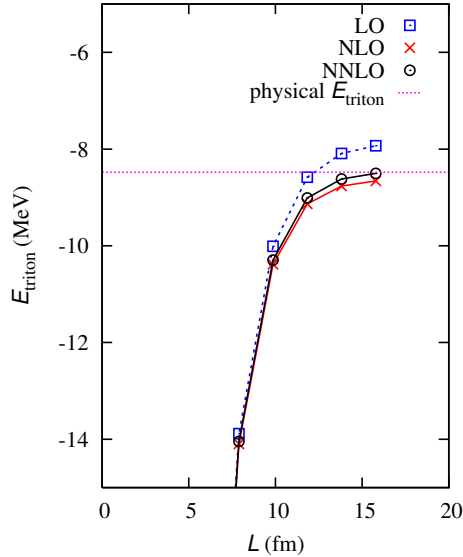}%
\caption{Triton energy versus periodic lattice length at LO, NLO, and NNLO.
\ For the NNLO results we show data for $c_{D}=1.0$ fitted to the physical
triton energy.}%
\label{triton_energy}%
\end{center}
\end{figure}

With the lattice transfer matrices $M_{\text{LO}}$, $M_{\text{NLO}}$,
$M_{\text{NNLO}}$, we use iterative sparse-matrix eigenvector methods to
compute the triton energy for cubic periodic lattices. \ We consider cubes
with side lengths $L=3,4,5,6,7,8$ and extract the infinite volume limit using
the asymptotic parameterization \cite{Luscher:1985dn},%
\begin{equation}
E_{\text{triton}}(L)\approx E_{\text{triton}}-\frac{C}{L}e^{-L/L_{0}}%
\text{.}\label{finiteL}%
\end{equation}
$L_{0}$ is a length scale associated with the physical size of the triton
wavefunction. \ For the NNLO calculation we fix the coefficient $c_{E}$ as a
function of $c_{D}$ by matching the physical triton energy at infinite volume,
$-8.48$~MeV. \ This constraint produces the solid line shown in Fig.
\ref{ce_vs_cd}. \ In the same figure the dotted line shows results obtained by
fitting to the \textquotedblleft pseudo\textquotedblright\ triton energy,
$-8.68$~MeV. \ This pseudo energy is an estimate of the triton energy when
$nn$ interactions are replaced with $np$ interactions \cite{Epelbaum:2002vt}.
\ This adjustment takes into account the systematic error in our
isospin-symmetric calculations with two-nucleon interactions matched to $np$
phase shifts.%
We note the similarity between Fig.~\ref{ce_vs_cd} and other plots of $c_{E}$
versus $c_{D}$ found in Fig.~2 of Ref.~\cite{Epelbaum:2002vt} and Fig.~2 of
Ref.~\cite{Navratil:2007aj}. \ In our case the plots are exactly linear due to
our perturbative treatment of the three-nucleon interactions. \ In
Fig.~\ref{triton_energy} we show $E_{\text{triton}}(L)$ versus $L$ measured in
physical units at LO, NLO, and NNLO. \ For the NNLO calculation we show data
for $c_{D}=1.0$ fitted to the physical triton energy. \ We see that the NLO
and NNLO corrections each appear small enough to be treated using perturbation
theory.%

\subsection{Neutron-deuteron scattering}

L\"{u}scher's formula relates the energy levels for a two-body system in a
finite periodic cube to scattering phase shifts at infinite volume
\cite{Luscher:1986pf,Luscher:1991ux}. \ We use this method to calculate
neutron-deuteron scattering phase shifts in the spin-doublet and spin-quartet
channels. \ For cubic lattice lengths $L=4,5,6,7,8$ we measure the
three-nucleon energy levels relative to the threshold energy for a
non-interacting neutron and deuteron in the same volume.

\begin{figure}
[ptb]
\begin{center}
\includegraphics[
height=3.0277in,
width=2.9603in
]%
{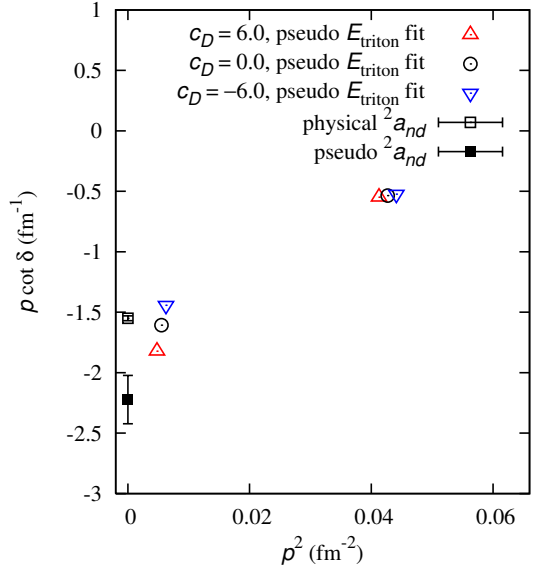}%
\caption{NNLO results for spin-doublet neutron-deuteron scattering. \ We plot
$p\cot\delta$ versus $p^{2}$ in the center-of-mass frame.}%
\label{scattering_length}%
\end{center}
\end{figure}

In Fig.~\ref{scattering_length} we plot $p\cot\delta$ versus $p^{2}$ in the
center-of-mass frame for the spin-doublet channel. \ Using the pseudo
$E_{\text{triton}}$ constraint for $c_{E}$, we show NNLO results for
$c_{D}=-6.0,$ $0.0,$ $6.0$. \ This can be compared with the physical
scattering length $^{2}a_{nd}=-0.645\pm0.003_{\text{exp.}}\pm0.007_{\text{th.}%
}$~fm \cite{Schoen:2003au} and pseudo scattering length $^{2}a_{nd}%
=-0.45(4)$~fm resulting from adjusting the strength of $nn$ interactions to
match $np$ interactions \cite{Epelbaum:2002vt}. \ A detailed discussion of
isospin-breaking contributions to the neutron-deuteron scattering lengths can
be found in Ref.~\cite{Witala:2003mr}. \ We match to the pseudo scattering
length and find that $c_{D}$ lies in the range from $-6.0$ to $+6.0$. \ This
is a rather loose constraint since we already expect $c_{D}\sim O(1)$ based on
the natural size of coefficients under renormalization group transformations.
\ We consider alternative methods for constraining $c_{D}$ later in our discussion.%

\begin{figure}
[ptb]
\begin{center}
\includegraphics[
height=2.7095in,
width=3.653in
]%
{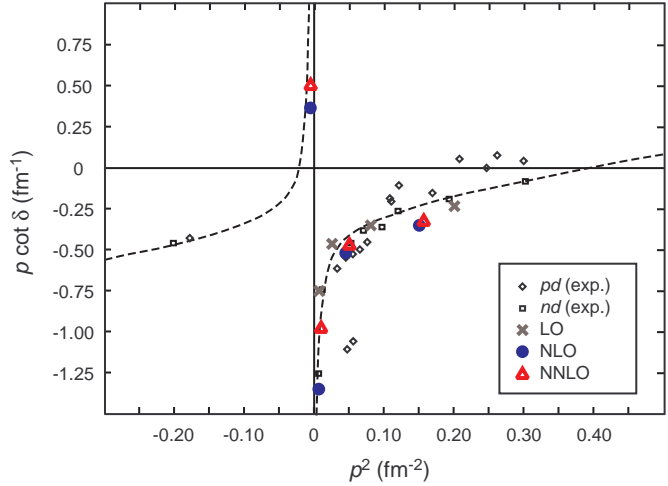}%
\caption{Plot of $p\cot\delta$ versus $p^{2}$ for the spin-doublet channel in
the center-of-mass frame. \ For the NNLO calculation we take $c_{D}=1.0$
fitted to the physical value for $E_{\text{triton}}$. \ The experimental
results are from Ref.~\cite{vanOers:1967}.}%
\label{nd_doublet}%
\end{center}
\end{figure}

In Fig.~\ref{nd_doublet} we plot $p\cot\delta$ for the spin-doublet channel
for a wider range of $p^{2}$. \ For the NNLO calculation we show data for
$c_{D}=1.0$ fitted to the physical value for $E_{\text{triton}}$. \ The
experimental results are $nd$ and $pd$ scattering data from the partial wave
analysis in Ref.~\cite{vanOers:1967}. \ The dashed line shows an empirical
model introduced in Ref.~\cite{vanOers:1967} with a pole singularity in
$p\cot\delta$ just below zero energy. \ This empirical model also accommodates
data points for the triton and $^{3}$He bound states at negative $p^{2}$. \ In
our lattice data we also find non-trivial scattering behavior just below zero
energy. \ The interpretation of these results and possible connections with
the Efimov effect at finite volume are currently being studied
\cite{Efimov:1971a,Efimov:1993a,Kreuzer:2008bi}. \ The deuteron break-up
threshold is near $p^{2}=0.07$~fm$^{-2}$, and the agreement between lattice
and experimental results for $nd$ scattering is quite good below break-up.
\ Above the break-up threshold our analysis using L\"{u}scher's finite volume
formula does not take into account mixing between $nd$ and $nnp$ three-nucleon
states. \ Therefore we expect significant errors in the case of strong mixing.%

\begin{figure}
[ptb]
\begin{center}
\includegraphics[
height=2.4907in,
width=3.2041in
]%
{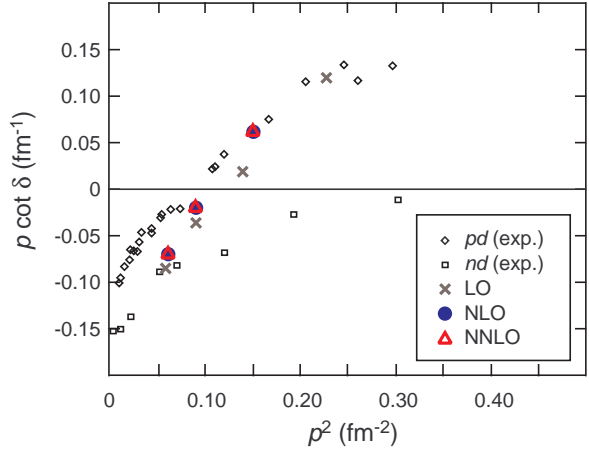}%
\caption{Plot of $p\cot\delta$ versus $p^{2}$ for the spin-quartet channel in
the center-of-mass frame. \ For the NNLO calculation we take $c_{D}=1.0$
fitted to the physical value for $E_{\text{triton}}$. \ The experimental
results are from Ref.~\cite{vanOers:1967}.}%
\label{nd_quartet}%
\end{center}
\end{figure}

In Fig.~\ref{nd_quartet} we plot $p\cot\delta$ for the spin-quartet channel.
\ Again we show experimental results for $nd$ and $pd$ scattering from
Ref.~\cite{vanOers:1967}, and for the NNLO calculation we present data for
$c_{D}=1.0$ fitted to the physical value for $E_{\text{triton}}$. \ The
lattice data agree with experimental results for $nd$ scattering below
deuteron break-up. \ However significant deviations appear above the break-up
threshold. \ This may indicate mixing effects between $nd$ and $nnp$ states.%

\section{Transfer matrices with auxiliary fields}

For systems with more than three nucleons, sparse-matrix calculations using
the lattice transfer matrix are not practical at large volume. \ Instead we
use projection Monte Carlo with auxiliary fields. \ A review of the
auxiliary-field formalism can be found in Ref.~\cite{Lee:2008fa}. \ We define
$M^{(n_{t})}(\pi_{I}^{\prime},s,s_{I})$ as the leading-order\ auxiliary-field
transfer matrix at time step $n_{t}$,%
\begin{align}
M^{(n_{t})}(\pi_{I}^{\prime},s,s_{I})  &  =\colon\exp\left\{  -H_{\text{free}%
}\alpha_{t}-\frac{g_{A}\alpha_{t}}{2f_{\pi}\sqrt{q_{\pi}}}%
{\displaystyle\sum_{\vec{n},S,I}}
\Delta_{S}\pi_{I}^{\prime}(\vec{n},n_{t})\rho_{S,I}^{a^{\dag},a}(\vec
{n})\right. \nonumber\\
&  \qquad\left.  +\sqrt{-C\alpha_{t}}\sum_{\vec{n}}s(\vec{n},n_{t}%
)\rho^{a^{\dag},a}(\vec{n})+i\sqrt{C_{I^{2}}\alpha_{t}}\sum_{\vec{n},I}%
s_{I}(\vec{n},n_{t})\rho_{I}^{a^{\dag},a}(\vec{n})\right\}  \colon.
\end{align}
We can write $M_{\text{LO}}$ as the normalized integral%
\begin{equation}
M_{\text{LO}}=\frac{%
{\displaystyle\int}
D\pi_{I}^{\prime}DsDs_{I}\;e^{-S_{\pi\pi}^{(n_{t})}-S_{ss}^{(n_{t})}}%
M^{(n_{t})}(\pi_{I}^{\prime},s,s_{I})}{%
{\displaystyle\int}
D\pi_{I}^{\prime}DsDs_{I}\;e^{-S_{\pi\pi}^{(n_{t})}-S_{ss}^{(n_{t})}}},
\label{LOaux}%
\end{equation}
where $S_{\pi\pi}^{(n_{t})}$ is the piece of the instantaneous pion action at
time step $n_{t}$,%
\begin{equation}
S_{\pi\pi}^{(n_{t})}(\pi_{I}^{\prime})=\frac{1}{2}\sum_{\vec{n},I}\pi
_{I}^{\prime}(\vec{n},n_{t})\pi_{I}^{\prime}(\vec{n},n_{t})-\frac{\alpha_{t}%
}{q_{\pi}}\sum_{\vec{n},I,l}\pi_{I}^{\prime}(\vec{n},n_{t})\pi_{I}^{\prime
}(\vec{n}+\hat{l},n_{t}),
\end{equation}
and $S_{ss}^{(n_{t})}$ is the auxiliary-field action at time step $n_{t}$,%
\begin{equation}
S_{ss}^{(n_{t})}=\frac{1}{2}\sum_{\vec{n},\vec{n}^{\prime}}s(\vec{n}%
,n_{t})f^{-1}(\vec{n}-\vec{n}^{\prime})s(\vec{n}^{\prime},n_{t})+\frac{1}%
{2}\sum_{I}\sum_{\vec{n},\vec{n}^{\prime}}s_{I}(\vec{n},n_{t})f^{-1}(\vec
{n}-\vec{n}^{\prime})s_{I}(\vec{n}^{\prime},n_{t}),
\end{equation}
with
\begin{equation}
f^{-1}(\vec{n}-\vec{n}^{\prime})=\frac{1}{L^{3}}\sum_{\vec{q}}\frac{1}%
{f(\vec{q})}e^{-i\vec{q}\cdot(\vec{n}-\vec{n}^{\prime})}\text{.}%
\end{equation}

The NLO and NNLO interactions are treated using perturbation theory. \ We let%
\begin{align}
U^{(n_{t})}(\varepsilon)  &  =\sum_{\vec{n}}\varepsilon_{\rho}(\vec{n}%
,n_{t})\rho^{a^{\dagger},a}(\vec{n})+\sum_{\vec{n},S}\varepsilon_{\rho_{S}%
}(\vec{n},n_{t})\rho_{S}^{a^{\dagger},a}(\vec{n})+\sum_{\vec{n},S}%
\varepsilon_{\Delta_{S}\rho}(\vec{n},n_{t})\Delta_{S}\rho^{a^{\dagger},a}%
(\vec{n})\nonumber\\
&  +\sum_{\vec{n},S,S^{\prime}}\varepsilon_{\Delta_{S}\rho_{S^{\prime}}}%
(\vec{n},n_{t})\Delta_{S}\rho_{S^{\prime}}^{a^{\dagger},a}(\vec{n})+\sum
_{\vec{n},l}\varepsilon_{\triangledown_{l}^{2}\rho}(\vec{n},n_{t}%
)\triangledown_{l}^{2}\rho^{a^{\dagger},a}(\vec{n})\nonumber\\
&  +\sum_{\vec{n},l,S}\varepsilon_{\triangledown_{l}^{2}\rho_{S}}(\vec
{n},n_{t})\triangledown_{l}^{2}\rho_{S}^{a^{\dagger},a}(\vec{n})+\sum_{\vec
{n},l}\varepsilon_{\Pi_{l}}(\vec{n},n_{t})\Pi_{l}^{a^{\dagger},a}(\vec
{n})+\sum_{\vec{n},l,S}\varepsilon_{\Pi_{l,S}}(\vec{n},n_{t})\Pi
_{l,S}^{a^{\dagger},a}(\vec{n}),
\end{align}
and%
\begin{align}
U_{I^{2}}^{(n_{t})}(\varepsilon)  &  =\sum_{\vec{n},I}\varepsilon_{\rho_{I}%
}(\vec{n},n_{t})\rho_{I}^{a^{\dagger},a}(\vec{n})+\sum_{\vec{n},S,I}%
\varepsilon_{\rho_{S,I}}(\vec{n},n_{t})\rho_{S,I}^{a^{\dagger},a}(\vec
{n})+\sum_{\vec{n},S,I}\varepsilon_{\Delta_{S}\rho_{I}}(\vec{n},n_{t}%
)\Delta_{S}\rho_{I}^{a^{\dagger},a}(\vec{n})\nonumber\\
&  +\sum_{\vec{n},S,S^{\prime},I}\varepsilon_{\Delta_{S}\rho_{S^{\prime},I}%
}(\vec{n},n_{t})\Delta_{S}\rho_{S^{\prime},I}^{a^{\dagger},a}(\vec{n}%
)+\sum_{\vec{n},l,I}\varepsilon_{\triangledown_{l}^{2}\rho_{I}}(\vec{n}%
,n_{t})\triangledown_{l}^{2}\rho_{I}^{a^{\dagger},a}(\vec{n})\nonumber\\
&  +\sum_{\vec{n},l,S,I}\varepsilon_{\triangledown_{l}^{2}\rho_{S,I}}(\vec
{n},n_{t})\triangledown_{l}^{2}\rho_{S,I}^{a^{\dagger},a}(\vec{n})+\sum
_{\vec{n},l,I}\varepsilon_{\Pi_{l,I}}(\vec{n},n_{t})\Pi_{l,I}^{a^{\dagger}%
,a}(\vec{n})+\sum_{\vec{n},l,S,I}\varepsilon_{\Pi_{l,S,I}}(\vec{n},n_{t}%
)\Pi_{l,S,I}^{a^{\dagger},a}(\vec{n}).
\end{align}
With these extra fields and linear functionals we define%
\begin{align}
&  M^{(n_{t})}(\pi_{I}^{\prime},s,s_{I},\varepsilon)\nonumber\\
&  =\colon\exp\left\{  -H_{\text{free}}\alpha_{t}-\frac{g_{A}\alpha_{t}%
}{2f_{\pi}\sqrt{q_{\pi}}}%
{\displaystyle\sum_{\vec{n},S,I}}
\Delta_{S}\pi_{I}^{\prime}(\vec{n},n_{t})\rho_{S,I}^{a^{\dag},a}(\vec
{n})\right. \nonumber\\
&  \left.  +\sqrt{-C\alpha_{t}}\sum_{\vec{n}}s(\vec{n},n_{t})\rho^{a^{\dag}%
,a}(\vec{n})+i\sqrt{C_{I^{2}}\alpha_{t}}\sum_{\vec{n},I}s_{I}(\vec{n}%
,n_{t})\rho_{I}^{a^{\dag},a}(\vec{n})+U^{(n_{t})}(\varepsilon)+U_{I^{2}%
}^{(n_{t})}(\varepsilon)\right\}  \colon.
\end{align}
In Ref.~\cite{Borasoy:2007vk,Lee:2008fa,Epelbaum:2008vj} there were factors of
$\sqrt{\alpha_{t}}$ multiplying $U^{(n_{t})}(\varepsilon)$ and $U_{I^{2}%
}^{(n_{t})}(\varepsilon)$. \ We have removed these factors here as they
complicate the discussion of the three-body interactions at NNLO. \ Let
$M^{(n_{t})}(\varepsilon)$ be the normalized integral,%
\begin{equation}
M^{(n_{t})}(\varepsilon)=\frac{%
{\displaystyle\int}
D\pi^{\prime}DsDs_{I}\;e^{-S_{\pi\pi}^{(n_{t})}-S_{ss}^{(n_{t})}}M^{(n_{t}%
)}(\pi_{I}^{\prime},s,s_{I},\varepsilon)}{%
{\displaystyle\int}
D\pi^{\prime}DsDs_{I}\;e^{-S_{\pi\pi}^{(n_{t})}-S_{ss}^{(n_{t})}}}.
\end{equation}
When all $\varepsilon$ fields are set to zero we recover $M_{\text{LO}}$,%
\begin{equation}
M^{(n_{t})}(0)=M_{\text{LO}}\text{.}%
\end{equation}

To first order in perturbation theory the NLO interactions in $M_{\text{NLO}}$
can be written as a sum of bilinear derivatives of $M^{(n_{t})}(\varepsilon)$
with respect to the $\varepsilon$ fields at $\varepsilon=0$,%
\begin{align}
M_{\text{NLO}}  &  =M_{\text{LO}}\nonumber\\
&  -\frac{1}{2}\Delta C\alpha_{t}\sum_{\vec{n}}\left.  \frac{\delta}%
{\delta\varepsilon_{\rho}(\vec{n},n_{t})}\frac{\delta}{\delta\varepsilon
_{\rho}(\vec{n},n_{t})}M^{(n_{t})}(\varepsilon)\right\vert _{\varepsilon
=0}\nonumber\\
&  +\frac{1}{2}C_{q^{2}}\alpha_{t}\sum_{\vec{n}}\left.  \frac{\delta}%
{\delta\varepsilon_{\rho}(\vec{n},n_{t})}\frac{\delta}{\delta\varepsilon
_{\triangledown_{l}^{2}\rho}(\vec{n},n_{t})}M^{(n_{t})}(\varepsilon
)\right\vert _{\varepsilon=0}+\;\cdots.
\end{align}
\bigskip Similarly for NNLO we have
\begin{equation}
M_{\text{NNLO}}=M_{\text{NLO}}+\frac{%
{\displaystyle\int}
D\pi^{\prime}DsDs_{I}\;e^{-S_{\pi\pi}^{(n_{t})}-S_{ss}^{(n_{t})}}\Delta
M_{\text{NNLO}}^{(n_{t})}\left(  \pi^{\prime}\right)  }{%
{\displaystyle\int}
D\pi^{\prime}DsDs_{I}\;e^{-S_{\pi\pi}^{(n_{t})}-S_{ss}^{(n_{t})}}},
\end{equation}
where%
\begin{align}
\Delta M_{\text{NNLO}}^{(n_{t})}\left(  \pi^{\prime}\right)   &
=M_{\text{contact}}^{(3N)(n_{t})}\left(  \pi^{\prime}\right)  +M_{\text{OPE}%
}^{(3N)(n_{t})}\left(  \pi^{\prime}\right) \nonumber\\
&  +M_{\text{TPE1}}^{(3N)(n_{t})}\left(  \pi^{\prime}\right)  +M_{\text{TPE2}%
}^{(3N)(n_{t})}\left(  \pi^{\prime}\right)  +M_{\text{TPE3}}^{(3N)(n_{t}%
)}\left(  \pi^{\prime}\right)  .
\end{align}
The three-nucleon contact interaction is%
\begin{equation}
M_{\text{contact}}^{(3N)(n_{t})}\left(  \pi^{\prime}\right)  =-\frac{1}%
{6}D_{\text{contact}}\alpha_{t}\sum_{\vec{n}}\left.  \left[  \frac{\delta
}{\delta\varepsilon_{\rho}(\vec{n},n_{t})}\right]  ^{3}M^{(n_{t})}%
(\varepsilon)\right\vert _{\varepsilon=0},
\end{equation}
and the one-pion exchange interaction has the form%
\begin{equation}
M_{\text{OPE}}^{(3N)(n_{t})}\left(  \pi^{\prime}\right)  =-D_{\text{OPE}}%
\frac{\alpha_{t}}{\sqrt{q_{\pi}}}\sum_{\vec{n},S,I}\left.  \Delta_{S}\pi
_{I}^{\prime}(\vec{n},n_{t})\frac{\delta}{\delta\varepsilon_{\rho_{S,I}}%
(\vec{n},n_{t})}\frac{\delta}{\delta\varepsilon_{\rho}(\vec{n},n_{t}%
)}M^{(n_{t})}(\varepsilon)\right\vert _{\varepsilon=0}\text{.}%
\end{equation}
The three two-pion exchange terms are%
\begin{align}
&  M_{\text{TPE1}}^{(3N)(n_{t})}\left(  \pi^{\prime}\right)  =-D_{\text{TPE1}%
}\frac{\alpha_{t}}{q_{\pi}}\nonumber\\
&  \times\sum_{\vec{n},S,I}\left[  \Delta_{S}\pi_{I}^{\prime}(\vec{n}%
,n_{t})\Delta_{S}\pi_{I}^{\prime}(\vec{n},n_{t})-\left\langle \Delta_{S}%
\pi_{I}^{\prime}(\vec{n},n_{t})\Delta_{S}\pi_{I}^{\prime}(\vec{n}%
,n_{t})\right\rangle \right]  \left.  \frac{\delta M^{(n_{t})}(\varepsilon
)}{\delta\varepsilon_{\rho}(\vec{n},n_{t})}\right\vert _{\varepsilon=0},
\end{align}%
\begin{align}
&  M_{\text{TPE2}}^{(3N)(n_{t})}\left(  \pi^{\prime}\right)  =-D_{\text{TPE2}%
}\frac{m_{\pi}^{2}\alpha_{t}}{q_{\pi}}\nonumber\\
&  \times\sum_{\vec{n},I}\left[  \square\pi_{I}^{\prime}(\vec{n},n_{t}%
)\square\pi_{I}^{\prime}(\vec{n},n_{t})-\left\langle \square\pi_{I}^{\prime
}(\vec{n},n_{t})\square\pi_{I}^{\prime}(\vec{n},n_{t})\right\rangle \right]
\left.  \frac{\delta M^{(n_{t})}(\varepsilon)}{\delta\varepsilon_{\rho}%
(\vec{n},n_{t})}\right\vert _{\varepsilon=0},
\end{align}%
\begin{align}
&  M_{\text{TPE3}}^{(3N)(n_{t})}\left(  \pi^{\prime}\right)  =-D_{\text{TPE3}%
}\frac{\alpha_{t}}{q_{\pi}}\nonumber\\
&  \times\sum_{\vec{n},S_{1},S_{2},S_{3}}\sum_{I_{1},I_{2},I_{3}}%
\varepsilon_{S_{1},S_{2},S_{3}}\varepsilon_{I_{1},I_{2},I_{3}}\Delta_{S_{1}%
}\pi_{I_{1}}^{\prime}(\vec{n},n_{t})\Delta_{S_{2}}\pi_{I_{2}}^{\prime}(\vec
{n},n_{t})\left.  \frac{\delta M^{(n_{t})}(\varepsilon)}{\delta\varepsilon
_{\rho_{S_{3},I_{3}}}(\vec{n},n_{t})}\right\vert _{\varepsilon=0}.
\end{align}

We extract the properties of the ground state using Euclidean-time projection.
\ Let $\left\vert \Psi^{\text{free}}\right\rangle $ be a Slater determinant of
free-particle standing waves on the lattice. \ We construct the trial state
$\left\vert \Psi(t^{\prime})\right\rangle $ using%
\begin{equation}
\left\vert \Psi(t^{\prime})\right\rangle =\left(  M_{\text{SU(4)}\not \pi
}\right)  ^{L_{t_{o}}}\left\vert \Psi^{\text{free}}\right\rangle ,
\label{L_t_o}%
\end{equation}
where $t^{\prime}=L_{t_{o}}\alpha_{t}$ and $L_{t_{o}}$ is the number of
\textquotedblleft outer\textquotedblright\ time steps. \ As the notation
suggests, the transfer matrix $M_{\text{SU(4)}\not \pi }$ is invariant under
an exact Wigner SU(4) symmetry and acts as an approximate low-energy filter.
\ The amplitude $Z(t)$ is defined as%
\begin{equation}
Z(t)=\left\langle \Psi(t^{\prime})\right\vert \left(  M_{\text{LO}}\right)
^{L_{t_{i}}}\left\vert \Psi(t^{\prime})\right\rangle , \label{L_t_i}%
\end{equation}
where $t=L_{t_{i}}\alpha_{t}$ and $L_{t_{i}}$ is the number of
\textquotedblleft inner\textquotedblright\ time steps. \ The transient energy%
\begin{equation}
E_{\text{LO}}(t+\alpha_{t}/2)
\end{equation}
is given by the ratio of the amplitudes for $t$ and $t+\alpha_{t}$,%
\begin{equation}
e^{-E_{\text{LO}}(t+\alpha_{t}/2)\cdot\alpha_{t}}=\frac{Z(t+\alpha_{t})}%
{Z(t)}.
\end{equation}
The ground state energy $E_{0,\text{LO}}$ equals the asymptotic limit,%
\begin{equation}
E_{0,\text{LO}}=\lim_{t\rightarrow\infty}E_{\text{LO}}(t+\alpha_{t}/2).
\end{equation}
We calculate these Euclidean-time projection amplitudes using auxiliary
fields. \ For a given configuration of auxiliary and pion fields, the
contribution to the amplitude $Z(t)$ is proportional to the determinant of an
$A\times A$ matrix of one-body amplitudes where $A$ is the number of nucleons.
\ Integrations over auxiliary and pion field configurations are computed using
hybrid Monte Carlo. \ Details of the method can be found in
Ref.~\cite{Lee:2005fk,Lee:2006hr,Borasoy:2006qn,Lee:2008fa}.

For the ground state energy at NLO and NNLO we compute expectation values of
$M_{\text{LO}}$, $M_{\text{NLO}}$, $M_{\text{NNLO}}$ inserted in the middle of
a string of $M_{\text{LO}}$ transfer matrices,%
\begin{equation}
Z_{M_{\text{LO}}}(t)=\left\langle \Psi(t^{\prime})\right\vert \left(
M_{\text{LO}}\right)  ^{L_{t_{i}}/2}M_{\text{LO}}\left(  M_{\text{LO}}\right)
^{L_{t_{i}}/2}\left\vert \Psi(t^{\prime})\right\rangle ,
\end{equation}%
\begin{equation}
Z_{M_{\text{NLO}}}(t)=\left\langle \Psi(t^{\prime})\right\vert \left(
M_{\text{LO}}\right)  ^{L_{t_{i}}/2}M_{\text{NLO}}\left(  M_{\text{LO}%
}\right)  ^{L_{t_{i}}/2}\left\vert \Psi(t^{\prime})\right\rangle ,
\end{equation}%
\begin{equation}
Z_{M_{\text{NNLO}}}(t)=\left\langle \Psi(t^{\prime})\right\vert \left(
M_{\text{LO}}\right)  ^{L_{t_{i}}/2}M_{\text{NNLO}}\left(  M_{\text{LO}%
}\right)  ^{L_{t_{i}}/2}\left\vert \Psi(t^{\prime})\right\rangle .
\end{equation}
From the ratio of amplitudes,%
\begin{equation}
\frac{Z_{M_{\text{NLO}}}(t)}{Z_{M_{\text{LO}}}(t)}=1-\Delta E_{\text{NLO}%
}(t)\alpha_{t}+\cdots,
\end{equation}
we define the transient NLO energy correction $\Delta E_{\text{NLO}}(t)$.
\ The ellipsis denotes terms which are beyond first order in the
NLO\ coefficients. \ The NLO ground state energy $E_{0,\text{NLO}}$ is
calculated using%
\begin{equation}
E_{0,\text{NLO}}=E_{0,\text{LO}}+\lim_{t\rightarrow\infty}\Delta
E_{\text{NLO}}(t).
\end{equation}
Similarly at NNLO we have%
\begin{equation}
\frac{Z_{M_{\text{NNLO}}}(t)}{Z_{M_{\text{LO}}}(t)}=1-\Delta E_{\text{NNLO}%
}(t)\alpha_{t}+\cdots,
\end{equation}
and
\begin{equation}
E_{0,\text{NNLO}}=E_{0,\text{NLO}}+\lim_{t\rightarrow\infty}\Delta
E_{\text{NNLO}}(t).
\end{equation}

\section{Precision tests of Monte Carlo simulations}

We use the three-nucleon system to test the auxiliary-field Monte Carlo
simulations. \ The same observables are calculated using both auxiliary-field
Monte Carlo and the exact transfer matrix without auxiliary fields. \ We
choose a small system so that stochastic errors are small enough to expose
disagreement at the $0.1\%-1\%$ level. \ We choose the spatial length of the
lattice to be $L=3$ and set the outer time steps $L_{t_{o}}=0$ and inner time
steps $L_{t_{i}}=4$. \ With $2048$ processors we generate a total of about
$10^{8}$ hybrid Monte Carlo trajectories. \ Each processor runs completely
independent trajectories, and we compute averages and stochastic errors by
comparing the results of all processors.

We choose $\left\vert \Psi^{\text{free}}\right\rangle $ to be a spin-doublet
isospin-doublet state built from the Slater determinant of standing waves
$\left\vert \psi_{1}\right\rangle $, $\left\vert \psi_{2}\right\rangle $,
$\left\vert \psi_{3}\right\rangle $ with%
\begin{equation}
\left\langle 0\right\vert a_{i,j}(\vec{n})\left\vert \psi_{1}\right\rangle
\propto\delta_{i,0}\delta_{j,1},\qquad\left\langle 0\right\vert a_{i,j}%
(\vec{n})\left\vert \psi_{2}\right\rangle \propto\delta_{i,1}\delta
_{j,1},\qquad\left\langle 0\right\vert a_{i,j}(\vec{n})\left\vert \psi
_{3}\right\rangle \propto\delta_{i,0}\delta_{j,0}\text{.}%
\end{equation}
In Table~\ref{precision_summary} we show Monte Carlo results for the energy
(MC) versus exact transfer matrix calculations (Exact) at LO, NLO, and NNLO.
\ The NNLO data uses $c_{D}=1.0$ with $c_{E}$ fitted to the physical triton
energy. \ In Table~\ref{precision_nlo} we compare Monte Carlo results with
exact transfer matrix calculations for the derivative of the energy with
respect to each NLO coefficient. \ In Table~\ref{precision_nnlo} we make the
same comparison for the derivative of the energy with respect to each NNLO
coefficient. \ The numbers in parentheses are the estimated stochastic errors.
\ In all cases the agreement between Monte Carlo results and exact transfer
calculations is consistent with estimated stochastic errors.

\begin{table}[tb]
\caption{Monte Carlo results for the energy versus exact transfer matrix
calculations}%
\label{precision_summary}
\begin{tabular}
[c]{||c|c|c||}\hline\hline
Energies & MC & Exact\\\hline
$E_{\text{LO}}(t+\alpha_{t}/2)$ [MeV] & $-48.873(18)$ & $-48.8823$\\\hline
$\Delta E_{\text{NLO}}(t)$ [MeV] & $0.5509(8)$ & $0.55100$\\\hline
$\Delta E_{\text{NNLO}}(t)$ [MeV] & $-0.967(3)$ & $-0.96718$\\\hline\hline
\end{tabular}
\end{table}

\begin{table}[tb]
\caption{Monte Carlo results versus exact transfer matrix calculations for the
derivative of the energy with respect to NLO coefficients}%
\label{precision_nlo}
\begin{tabular}
[c]{||c|c|c||}\hline\hline
NLO energy derivatives & MC & Exact\\\hline
$\frac{\partial\left(  \Delta E_{\text{NLO}}(t)\right)  }{\partial\left(
\Delta C\right)  }$ [$10^{4}$ MeV$^{3}$] & $3.9037(12)$ & $3.90226$\\\hline
$\frac{\partial\left(  \Delta E_{\text{NLO}}(t)\right)  }{\partial\left(
\Delta C_{I^{2}}\right)  }$ [$10^{4}$ MeV$^{3}$] & $-4.847(2)$ &
$-4.84331$\\\hline
$\frac{\partial\left(  \Delta E_{\text{NLO}}(t)\right)  }{\partial\left(
C_{q^{2}}\right)  }$ [$10^{9}$ MeV$^{5}$] & $-2.0105(6)$ & $-2.01059$\\\hline
$\frac{\partial\left(  \Delta E_{\text{NLO}}(t)\right)  }{\partial\left(
C_{I^{2},q^{2}}\right)  }$ [$10^{9}$ MeV$^{5}$] & $2.9230(14)$ &
$2.92424$\\\hline
$\frac{\partial\left(  \Delta E_{\text{NLO}}(t)\right)  }{\partial\left(
C_{S^{2},q^{2}}\right)  }$ [$10^{9}$ MeV$^{5}$] & $0.1860(12)$ &
$0.18411$\\\hline
$\frac{\partial\left(  \Delta E_{\text{NLO}}(t)\right)  }{\partial\left(
C_{S^{2},I^{2},q^{2}}\right)  }$ [$10^{9}$ MeV$^{5}$] & $5.094(2)$ &
$5.09371$\\\hline
$\frac{\partial\left(  \Delta E_{\text{NLO}_{3}}(t)\right)  }{\partial\left(
C_{(q\cdot S)^{2}}\right)  }$ [$10^{9}$ MeV$^{5}$] & $-1.5892(3)$ &
$-1.58898$\\\hline
$\frac{\partial\left(  \Delta E_{\text{NLO}}(t)\right)  }{\partial\left(
C_{I^{2},(q\cdot S)^{2}}\right)  }$ [$10^{9}$ MeV$^{5}$] & $6.8019(11)$ &
$6.80197$\\\hline
$\frac{\partial\left(  \Delta E_{\text{NLO}}(t)\right)  }{\partial\left(
C_{(iq\times S)\cdot k}\right)  }$ [$10^{9}$ MeV$^{5}$] & $0.3417(2)$ &
$0.34164$\\\hline
$\frac{\partial\left(  \Delta E_{\text{NLO}}(t)\right)  }{\partial\left(
C_{I^{2},(iq\times S)\cdot k}\right)  }$ [$10^{9}$ MeV$^{5}$] & $-1.0092(5)$ &
$-1.00932$\\\hline\hline
\end{tabular}
\end{table}

\begin{table}[tb]
\caption{Monte Carlo results versus exact transfer matrix calculations for the
derivative of the energy with respect to NNLO coefficients}%
\label{precision_nnlo}
\begin{tabular}
[c]{||c|c|c||}\hline\hline
NNLO energy derivatives & MC & Exact\\\hline
$\frac{\partial\left(  \Delta E_{\text{NNLO}}(t)\right)  }{\partial\left(
D_{\text{contact}}\right)  }$ [$10^{8}$ MeV$^{6}$] & $1.162(4)$ &
$1.1609$\\\hline
$\frac{\partial\left(  \Delta E_{\text{NNLO}}(t)\right)  }{\partial\left(
D_{\text{OPE}}\right)  }$ [$10^{7}$ MeV$^{5}$] & $-5.858(6)$ & $-5.8623$%
\\\hline
$\frac{\partial\left(  \Delta E_{\text{NNLO}}(t)\right)  }{\partial\left(
D_{\text{TPE1}}\right)  }$ [$10^{5}$ MeV$^{4}$] & $14.46(5)$ & $14.468$%
\\\hline
$\frac{\partial\left(  \Delta E_{\text{NNLO}}(t)\right)  }{\partial\left(
D_{\text{TPE2}}\right)  }$ [$10^{5}$ MeV$^{4}$] & $2.24(3)$ & $2.2595$\\\hline
$\frac{\partial\left(  \Delta E_{\text{NNLO}}(t)\right)  }{\partial\left(
D_{\text{TPE3}}\right)  }$ [$10^{5}$ MeV$^{4}$] & $-10.02(6)$ & $-10.022$%
\\\hline\hline
\end{tabular}
\end{table}

\section{Energy of the $\alpha$ particle}

We simulate the $\alpha$ particle on cubic periodic lattices with length
$L=5,6,7,8$. \ These correspond with physical lengths $L=9.9$, $11.8$, $13.8$,
$15.8$~fm. \ For $\left\vert \Psi^{\text{free}}\right\rangle $ we take the
Slater determinant formed by standing waves%
\begin{equation}
\left\langle 0\right\vert a_{i,j}(\vec{n})\left\vert \psi_{1}\right\rangle
\propto\delta_{i,0}\delta_{j,1},\qquad\left\langle 0\right\vert a_{i,j}%
(\vec{n})\left\vert \psi_{2}\right\rangle \propto\delta_{i,1}\delta_{j,1},
\end{equation}%
\begin{equation}
\left\langle 0\right\vert a_{i,j}(\vec{n})\left\vert \psi_{3}\right\rangle
\propto\delta_{i,0}\delta_{j,0},\qquad\left\langle 0\right\vert a_{i,j}%
(\vec{n})\left\vert \psi_{4}\right\rangle \propto\delta_{i,1}\delta
_{j,0}\text{.}%
\end{equation}
\newline For each value of $L_{t_{i}}$ a total of about $6\times10^{6}$ hybrid
Monte Carlo trajectories are generated by $2048$ processors, each running
completely independent trajectories. \ Averages and stochastic errors are
computed by comparing the results of all processors.

\begin{figure}
[ptb]
\begin{center}
\includegraphics[
height=5.444in,
width=4.6086in
]%
{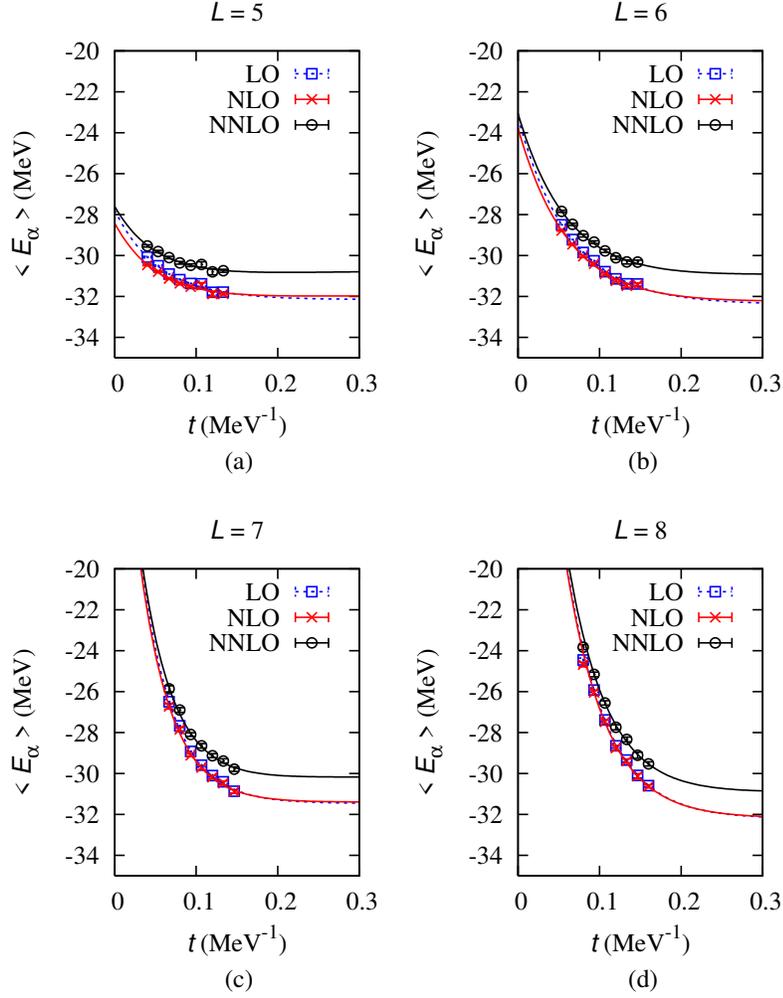}%
\caption{Monte Carlo results for $\alpha$-particle energies versus Euclidean
time $t$ at LO, NLO, and NNLO. \ We also plot fitted asymptotic expressions.
\ The NNLO data uses $c_{D}=1.0$ with $c_{E}$ fitted to the physical triton
energy.}%
\label{alpha_lt_all}%
\end{center}
\end{figure}

For $L=5,6,7,8$ we show Monte Carlo results in Fig.~\ref{alpha_lt_all} for%
\begin{equation}
E_{\text{LO}}(t),\quad E_{\text{LO}}(t)+\Delta E_{\text{NLO}}(t),\quad
E_{\text{LO}}(t)+\Delta E_{\text{NLO}}(t)+\Delta E_{\text{NNLO}}(t),
\label{energies_v_time}%
\end{equation}
versus Euclidean time $t$. \ These are labelled as LO, NLO, and NNLO
respectively. \ The NNLO data uses $c_{D}=1.0$ with $c_{E}$ fitted to the
physical triton energy. \ In addition to the Monte Carlo data we plot the
asymptotic forms for each term in Eq.~(\ref{energies_v_time}) using%
\begin{equation}
E_{\text{LO}}(t)\approx E_{0,\text{LO}}+A_{\text{LO}}e^{-\delta E\cdot t},
\label{asymptotic1}%
\end{equation}%
\begin{equation}
\Delta E_{\text{NLO}}(t)\approx E_{0,\text{NLO}}-E_{0,\text{LO}}%
+B_{\text{NLO}}e^{-\delta E\cdot t/2}, \label{asymptotic2}%
\end{equation}%
\begin{equation}
\Delta E_{\text{NNLO}}(t)\approx E_{0,\text{NNLO}}-E_{0,\text{NLO}%
}+B_{\text{NNLO}}e^{-\delta E\cdot t/2}. \label{asymptotic3}%
\end{equation}
The unknown coefficients $A_{\text{LO}}$, $B_{\text{NLO}}$, $B_{\text{NNLO}}$,
and energy gap $\delta E$, are determined by least squares fitting. \ The
$e^{-\delta E\cdot t}$ dependence in Eq.~(\ref{asymptotic1}) comes from the
contribution of low-energy excitations with energy gap $\delta E$ above the
$\alpha$ particle. \ The $e^{-\delta E\cdot t/2}$ dependence in
Eq.~(\ref{asymptotic2}) is due to matrix elements of $M_{\text{NLO}}$ between
the $\alpha$ particle and low-energy excitations at $\delta E$. \ Similarly
the $e^{-\delta E\cdot t/2}$ dependence in Eq.~(\ref{asymptotic3}) is from
matrix elements of $M_{\text{NNLO}}$. \ The chi-squared per degree of freedom
for the fits are $1.5$ for $L=5,$ $1.0$ for $L=6$, $0.7$ for $L=7$, and $0.6$
for $L=8$. \ The small relative size of the NLO\ and NNLO energy corrections
suggest that the perturbative treatment of these terms appears reasonable.%

%

\begin{figure}
[ptb]
\begin{center}
\includegraphics[
height=5.6948in,
width=4.6086in
]%
{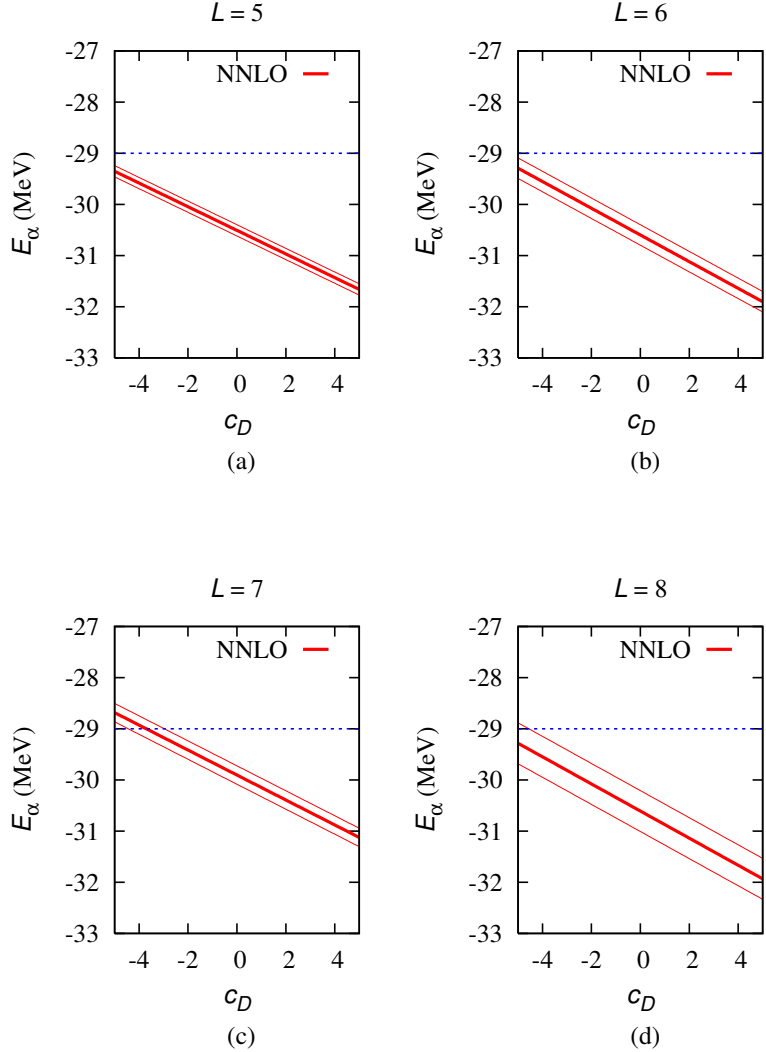}%
\caption{Energy of the $\alpha$ particle at NNLO versus $c_{D}$. \ The contact
interaction $c_{E}$ is fitted to the physical triton energy. \ The dotted line
is the estimated Coulomb-subtracted energy $-29.0$~MeV.}%
\label{alpha_cd}%
\end{center}
\end{figure}

In Fig.~\ref{alpha_cd} we plot the NNLO $\alpha$-particle energy versus
$c_{D}$, with $c_{E}$ fitted to the physical triton energy. \ The bands
indicate the estimated error due to stochastic noise and asymptotic fits at
large $t$. \ The $\alpha$ energy should approach the infinite volume limit
from below, similar to our results for the triton energy. \ Hence the
deviation between data at $L=7$ and $L=8$ is likely due to stochastic noise
and fit errors rather than finite volume effects. \ The $\alpha$ energy shown
at $-29.0$~MeV is the estimated Coulomb-subtracted energy
\cite{Pudliner:1997ck}. \ At large volumes the best agreement with the
Coulomb-subtracted $\alpha$ energy occurs at $c_{D}\approx-4$. \ The $\alpha$
binding increases in strength by $0.2$~MeV for each unit increase in $c_{D}$,
and so we find reasonable agreement for all values of $c_{D}\sim O(1)$. \ If
instead we fit $c_{E}$ according to the pseudo triton energy, then the lines
in Fig.~\ref{alpha_cd} shift downward in energy by about $2$~MeV. \ The pseudo
$\alpha$ energy with $pp$- and $nn$-forces matched to $np$-forces is estimated
to be $-29.8(1)$~MeV \cite{Epelbaum:2002vt}.

\section{Summary and discussion}

We have presented the first study of low-energy few-nucleon systems on the
lattice at next-to-next-to-leading order in chiral effective field theory.
\ We computed nucleon-nucleon phase shifts and $S$-$D$ mixing angle on the
lattice and used scattering data to determine unknown two-nucleon operator
coefficients. \ In the three-nucleon system we calculated the triton energy
and determined neutron-deuteron phase shifts using L\"{u}scher's finite volume
method. \ These were used to constrain the two cutoff-dependent three-body
coefficients, $c_{D}$ and $c_{E}$. \ For the four-nucleon system we recast the
lattice action in terms of auxiliary fields and used projection Monte Carlo to
calculate the energy of the $\alpha$ particle.

There are several ways in which the lattice calculations presented here can be
improved and extended in future work. \ One improvement is the inclusion of
isospin-breaking effects due to Coulomb interactions and quark mass
differences. \ The framework for isospin-violating effects in chiral effective
field theory has been developed over the past decade
\cite{vanKolck:1996rm,vanKolck:1997fu,Friar:1999zr,Epelbaum:1999zn,Walzl:2000cx,Friar:2003yv,Friar:2004ca,Epelbaum:2004xf,Epelbaum:2005fd}%
. \ Work is currently underway to implement these effects within the lattice formalism.

Another area of improvement concerns the $P$-wave phase shifts for our
leading-order lattice action. \ NLO corrections to the phase shifts are
substantial for nucleon momenta above $100$~MeV. \ This seems not to cause any
problems for the light $S$-shell nuclei considered here. \ However for
$P$-shell nuclei we may find corrections\ strong enough to spoil the
perturbative treatment of higher-order effects. \ In
Ref.~\cite{Epelbaum:2008vj} this problem has already been resolved in neutron
matter calculations using a new leading-order action LO$_{3}$. \ This lattice
action uses spin-isospin projection operators to produce Gaussian smearing
only in even partial wave channels. \ The implementation of the LO$_{3}$
action in Monte Carlo simulations with both protons and neutrons is
computationally more intensive than the pure neutron simulations in
Ref.~\cite{Epelbaum:2008vj}. \ The auxiliary-field formalism requires a total
of $16$ auxiliary fields and some increase in the sign/phase cancellations
relative to the LO$_{2}$ action. \ Nevertheless studies of light $P$-shell
nuclei using the LO$_{3}$ action are planned in the near future.

One recent paper constrains the cutoff-dependent coefficient $c_{D}$ from the
triton beta decay rate \cite{Gazit:2008ma}. \ From the point of view of
computing the spectrum of light nuclei, the easiest method for determining
$c_{D}$ is by means of the $\alpha$-particle energy.\ \ This has the added
benefit of removing systematic errors from the four nucleon system. \ If
however we also want to accurately describe the chiral interactions of
nucleons and light nuclei with soft pions, then it would be best to measure
$c_{D}$ directly from the SU(2) axial coupling to two nucleon states. \ This
is studied in Ref.~\cite{Hanhart:2000gp} using pion production data in $pp$
scattering. \ Unfortunately the pion production energy threshold is too high
to be accessible at our lattice spacing and extrapolations are required. \ In
the future another approach may be possible using direct theory-to-theory
matching. \ In this technique lattice QCD would be used to calculate the axial
charge of two-neutron scattering states in a periodic cube. \ This calculated
value of the axial charge could then be used to fix $c_{D}$ for any chosen
lattice spacing in lattice chiral effective field theory.

\section{Acknowledgements}

We are grateful for discussions with Hans Hammer and Simon Kreuzer. \ Partial
financial support from the Deutsche Forschungsgemeinschaft (SFB/TR 16),
Helmholtz Association (contract number VH-NG-222 and VH-VI-231), and U.S.
Department of Energy (DE-FG02-03ER41260) are acknowledged. \ This work was
further supported by the EU HadronPhysics2 project \textquotedblleft Study of
strongly interacting matter\textquotedblright. \ The computational resources
for this project were provided by the J\"{u}lich Supercomputing Centre at the
Forschungszentrum J\"{u}lich.

\appendix*

\section{Lattice notation}

The vector $\vec{n}$ represents integer-valued lattice vectors on a
three-dimensional spatial lattice, and $\vec{p},$ $\vec{q},$ $\vec{k}$
represent integer-valued momentum lattice vectors.$\ \ \hat{l}=\hat{1}$,
$\hat{2}$, $\hat{3}$ are unit lattice vectors in the spatial directions, $a$
is the spatial lattice spacing, and $L$ is the length of the cubic spatial
lattice in each direction. \ The lattice time step is $a_{t}$, and $n_{t}$
labels the number of time steps. \ We define $\alpha_{t}$ as the ratio between
lattice spacings, $\alpha_{t}=a_{t}/a$. \ Throughout our lattice discussion we
use dimensionless parameters and operators, which correspond with physical
values multiplied by the appropriate power of $a$. \ Final results are
presented in physical units with the corresponding unit stated explicitly.

We use $a$ and $a^{\dagger}$ to denote annihilation and creation operators.
\ We make explicit all spin and isospin indices,%
\begin{align}
a_{0,0}  &  =a_{\uparrow,p},\text{ \ }a_{0,1}=a_{\uparrow,n},\\
a_{1,0}  &  =a_{\downarrow,p},\text{ \ }a_{1,1}=a_{\downarrow,n}.
\end{align}
The first subscript is for spin and the second subscript is for isospin. \ We
use $\tau_{I}$ with $I=1,2,3$ to represent Pauli matrices acting in isospin
space and $\sigma_{S}$ with $S=1,2,3$ to represent Pauli matrices acting in
spin space.

We use the eight vertices of a unit cube on the lattice to define spatial
derivatives. \ For each spatial direction $l=1,2,3$ and any lattice function
$f(\vec{n})$, let%
\begin{equation}
\Delta_{l}f(\vec{n})=\frac{1}{4}\sum_{\substack{\nu_{1},\nu_{2},\nu_{3}%
=0,1}}(-1)^{\nu_{l}+1}f(\vec{n}+\vec{\nu}),\qquad\vec{\nu}=\nu_{1}\hat{1}%
+\nu_{2}\hat{2}+\nu_{3}\hat{3}. \label{derivative}%
\end{equation}
We also define the double spatial derivative along direction $l$,%
\begin{equation}
\triangledown_{l}^{2}f(\vec{n})=f(\vec{n}+\hat{l})+f(\vec{n}-\hat{l}%
)-2f(\vec{n}).
\end{equation}
For the three-body NNLO interactions we also use the notation%
\begin{equation}
\square f(\vec{n})=\frac{1}{8}\sum_{\substack{\nu_{1},\nu_{2},\nu_{3}%
=0,1}}f(\vec{n}+\vec{\nu}),\qquad\vec{\nu}=\nu_{1}\hat{1}+\nu_{2}\hat{2}%
+\nu_{3}\hat{3}.
\end{equation}

\subsection{Local densities and currents}

We define the local density,%
\begin{equation}
\rho^{a^{\dagger},a}(\vec{n})=\sum_{i,j=0,1}a_{i,j}^{\dagger}(\vec{n}%
)a_{i,j}(\vec{n}),
\end{equation}
which is invariant under Wigner's SU(4) symmetry \cite{Wigner:1937}.
\ Similarly we define the local spin density for $S=1,2,3,$%
\begin{equation}
\rho_{S}^{a^{\dagger},a}(\vec{n})=\sum_{i,j,i^{\prime}=0,1}a_{i,j}^{\dagger
}(\vec{n})\left[  \sigma_{S}\right]  _{ii^{\prime}}a_{i^{\prime},j}(\vec{n}),
\end{equation}
isospin density for $I=1,2,3,$%
\begin{equation}
\rho_{I}^{a^{\dagger},a}(\vec{n})=\sum_{i,j,j^{\prime}=0,1}a_{i,j}^{\dagger
}(\vec{n})\left[  \tau_{I}\right]  _{jj^{\prime}}a_{i,j^{\prime}}(\vec{n}),
\end{equation}
and spin-isospin density for $S,I=1,2,3,$%
\begin{equation}
\rho_{S,I}^{a^{\dagger},a}(\vec{n})=\sum_{i,j,i^{\prime},j^{\prime}%
=0,1}a_{i,j}^{\dagger}(\vec{n})\left[  \sigma_{S}\right]  _{ii^{\prime}%
}\left[  \tau_{I}\right]  _{jj^{\prime}}a_{i^{\prime},j^{\prime}}(\vec{n}).
\end{equation}

For each static density we also have an associated current density. \ Similar
to the definition of the lattice derivative $\Delta_{l}$ in
Eq.~(\ref{derivative}), we use the eight vertices of a unit cube,
\begin{equation}
\vec{\nu}=\nu_{1}\hat{1}+\nu_{2}\hat{2}+\nu_{3}\hat{3},
\end{equation}
for $\nu_{1},\nu_{2},\nu_{3}=0,1$. \ Let $\vec{\nu}(-l)$ for $l=1,2,3$ be the
result of reflecting the $l^{\text{th}}$-component of $\vec{\nu}$ about the
center of the cube,%
\begin{equation}
\vec{\nu}(-l)=\vec{\nu}+(1-2\nu_{l})\hat{l}.
\end{equation}
Omitting factors of $i$ and $1/m$, we can write the $l^{\text{th}}$-component
of the SU(4)-invariant current density as%
\begin{equation}
\Pi_{l}^{a^{\dagger},a}(\vec{n})=\frac{1}{4}\sum_{\substack{\nu_{1},\nu
_{2},\nu_{3}=0,1}}\sum_{i,j=0,1}(-1)^{\nu_{l}+1}a_{i,j}^{\dagger}(\vec{n}%
+\vec{\nu}(-l))a_{i,j}(\vec{n}+\vec{\nu}).
\end{equation}
Similarly the $l^{\text{th}}$-component of spin current density is%
\begin{equation}
\Pi_{l,S}^{a^{\dagger},a}(\vec{n})=\frac{1}{4}\sum_{\substack{\nu_{1},\nu
_{2},\nu_{3}=0,1}}\sum_{i,j,i^{\prime}=0,1}(-1)^{\nu_{l}+1}a_{i,j}^{\dagger
}(\vec{n}+\vec{\nu}(-l))\left[  \sigma_{S}\right]  _{ii^{\prime}}a_{i^{\prime
},j}(\vec{n}+\vec{\nu}),
\end{equation}
$l^{\text{th}}$-component of isospin current density is%
\begin{equation}
\Pi_{l,I}^{a^{\dagger},a}(\vec{n})=\frac{1}{4}\sum_{\substack{\nu_{1},\nu
_{2},\nu_{3}=0,1}}\sum_{i,j,j^{\prime}=0,1}(-1)^{\nu_{l}+1}a_{i,j}^{\dagger
}(\vec{n}+\vec{\nu}(-l))\left[  \tau_{I}\right]  _{jj^{\prime}}a_{i,j^{\prime
}}(\vec{n}+\vec{\nu}),
\end{equation}
and $l^{\text{th}}$-component of spin-isospin current density is%
\begin{equation}
\Pi_{l,S,I}^{a^{\dagger},a}(\vec{n})=\frac{1}{4}\sum_{\substack{\nu_{1}%
,\nu_{2},\nu_{3}=0,1}}\sum_{i,j,i^{\prime},j^{\prime}=0,1}(-1)^{\nu_{l}%
+1}a_{i,j}^{\dagger}(\vec{n}+\vec{\nu}(-l))\left[  \sigma_{S}\right]
_{ii^{\prime}}\left[  \tau_{I}\right]  _{jj^{\prime}}a_{i^{\prime},j^{\prime}%
}(\vec{n}+\vec{\nu}).
\end{equation}

\subsection{Instantaneous free pion action}

The lattice action for free pions with purely instantaneous propagation is%
\begin{equation}
S_{\pi\pi}(\pi_{I})=\alpha_{t}(\tfrac{m_{\pi}^{2}}{2}+3)\sum_{\vec{n},n_{t}%
,I}\pi_{I}(\vec{n},n_{t})\pi_{I}(\vec{n},n_{t})-\alpha_{t}\sum_{\vec{n}%
,n_{t},I,l}\pi_{I}(\vec{n},n_{t})\pi_{I}(\vec{n}+\hat{l},n_{t}),
\end{equation}
where $\pi_{I}$ is the pion field labelled with isospin index $I$. \ It is
convenient to define a rescaled pion field, $\pi_{I}^{\prime}$,%
\begin{equation}
\pi_{I}^{\prime}(\vec{n},n_{t})=\sqrt{q_{\pi}}\pi_{I}(\vec{n},n_{t}),
\end{equation}%
\begin{equation}
q_{\pi}=\alpha_{t}(m_{\pi}^{2}+6).
\end{equation}
Then%
\begin{equation}
S_{\pi\pi}(\pi_{I}^{\prime})=\frac{1}{2}\sum_{\vec{n},n_{t},I}\pi_{I}^{\prime
}(\vec{n},n_{t})\pi_{I}^{\prime}(\vec{n},n_{t})-\frac{\alpha_{t}}{q_{\pi}}%
\sum_{\vec{n},n_{t},I,l}\pi_{I}^{\prime}(\vec{n},n_{t})\pi_{I}^{\prime}%
(\vec{n}+\hat{l},n_{t}). \label{pionaction}%
\end{equation}

In momentum space the action is%
\begin{equation}
S_{\pi\pi}(\pi_{I}^{\prime})=\frac{1}{L^{3}}\sum_{I,\vec{k}}\pi_{I}^{\prime
}(-\vec{k},n_{t})\pi_{I}^{\prime}(\vec{k},n_{t})\left[  \frac{1}{2}%
-\frac{\alpha_{t}}{q_{\pi}}\sum_{l}\cos k_{l}\right]  .
\end{equation}
The instantaneous pion correlation function at spatial separation $\vec{n}$ is%
\begin{align}
\left\langle \pi_{I}^{\prime}(\vec{n},n_{t})\pi_{I}^{\prime}(\vec{0}%
,n_{t})\right\rangle  &  =\frac{\int D\pi_{I}^{\prime}\;\pi_{I}^{\prime}%
(\vec{n},n_{t})\pi_{I}^{\prime}(\vec{0},n_{t})\;\exp\left[  -S_{\pi\pi
}\right]  }{\int D\pi_{I}^{\prime}\;\exp\left[  -S_{\pi\pi}\right]  }\text{
\ (no sum on }I\text{)}\nonumber\\
&  =\frac{1}{L^{3}}\sum_{\vec{k}}e^{-i\vec{k}\cdot\vec{n}}D_{\pi}(\vec{k}),
\end{align}
where%
\begin{equation}
D_{\pi}(\vec{k})=\frac{1}{1-\tfrac{2\alpha_{t}}{q_{\pi}}\sum_{l}\cos k_{l}}.
\end{equation}
It is also useful to define the two-derivative pion correlator, $G_{S_{1}%
S_{2}}(\vec{n})$,%
\begin{align}
G_{S_{1}S_{2}}(\vec{n})  &  =\left\langle \Delta_{S_{1}}\pi_{I}^{\prime}%
(\vec{n},n_{t})\Delta_{S_{2}}\pi_{I}^{\prime}(\vec{0},n_{t})\right\rangle
\text{ \ (no sum on }I\text{)}\nonumber\\
&  =\frac{1}{16}\sum_{\nu_{1},\nu_{2},\nu_{3}=0,1}\sum_{\nu_{1}^{\prime}%
,\nu_{2}^{\prime},\nu_{3}^{\prime}=0,1}(-1)^{\nu_{S_{1}}}(-1)^{\nu_{S_{2}%
}^{\prime}}\left\langle \pi_{I}^{\prime}(\vec{n}+\vec{\nu}-\vec{\nu}^{\prime
},n_{t})\pi_{I}^{\prime}(\vec{0},n_{t})\right\rangle .
\end{align}

\bibliographystyle{apsrev}
\bibliography{NuclearMatter}

\end{document}